\documentclass[traditabstract]{aa}

\pdfoutput=1
\usepackage{graphicx}
\usepackage[font=small]{caption}
\usepackage{indentfirst}
\usepackage{tabu}
\usepackage[percent]{overpic}
\usepackage{transparent}
\usepackage{ragged2e}
\usepackage{longtable}

\usepackage{natbib}
\usepackage{booktabs}
\usepackage{fixltx2e}

\usepackage{amssymb}
\usepackage{txfonts}
\usepackage[a]{esvect}
\usepackage{tikz}

\newcommand{\degree}[0]{$^{\circ}$}

\newcommand{\cbra}[1]{\left( #1 \right)}      % put the argument between parentheses (curve brackets)
\newcommand{\avg}[1]{\left< #1 \right>}         % average
\let\baraccent=\=
\renewcommand{\=}[1]{\stackrel{#1}{=}} % for putting numbers above =
\let\arrowaccent=\>
\renewcommand{\>}[1]{\stackrel{#1}{\Rightarrow}} % for putting numbers above =>

\makeatletter
\newcommand{\rmnum}[1]{{\footnotesize{\expandafter\@slowromancap\romannumeral #1@}}}
\newcommand{\Rmnum}[1]{{\expandafter\@slowromancap\romannumeral #1@}}
\makeatother

\everymath={\rm}

\begin{document}

\title{Inefficient jet-induced star formation in Centaurus A:\\ \Large{High resolution ALMA observations of the northern filaments}\thanks{This paper makes use of the following ALMA data: ADS/JAO.ALMA$\#$2015.1.01019.S.}$^,$\thanks{Table \ref{table:specCO} and a catalogue of the molecular clouds are only available at the CDS via anonymous ftp to cdsarc.u-strasbg.fr or via http://cdsarc.u-strasbg.fr/viz-bin/qcat?J/A+A/XXX/XXX}}

\author{
   Q. Salom\'e\inst{1}  \and
   P. Salom\'e\inst{2}  \and
   M.-A. Miville-Desch\^enes\inst{1} \and
   F. Combes\inst{2,3}  \and
   S. Hamer\inst{4}
}

\institute{
   Institut d’Astrophysique Spatiale, CNRS, Univ. Paris-Sud, Université Paris-Saclay, bâtiment 121, 91405 Orsay cedex, France \\ email: quentin.salome@ias.u-psud.fr \and
   LERMA, Observatoire de Paris, PSL Research Univ., CNRS, Sorbonne Univ., UPMC Univ. Paris 06, 75014 Paris, France \and
   Coll\`ege de France, 11 place Marcelin Berthelot, 75005 Paris \and
   CRAL, Observatoire de Lyon, CNRS, Universit\'e Lyon 1, 9 Avenue Ch. Andr\'e, 69561 Saint Genis Laval cedex, France
}

\date{Received 23 June 2017 / Accepted ??}

\titlerunning{The northern filaments of Centaurus A with ALMA}
\authorrunning{Salomé et al.}

\abstract{
   NGC 5128 (Centaurus A) is one of the best targets to study AGN feedback in the local Universe. At 13.5 kpc from the galaxy, optical filaments with recent star formation lie along the radio jet direction. This region is a testbed for positive feedback, here through jet-induced star formation. Atacama Pathfinder EXperiment (APEX) observations have revealed strong CO emission in star-forming regions and in regions with no detected tracers of star formation activity. In cases where star formation is observed, this activity appears to be inefficient compared to the Kennicutt-Schmidt relation.
   We used the Atacama Large Millimeter/submillimeter Array (ALMA) to map the $^{12}CO$(1-0) emission all along the filaments of NGC 5128 at a resolution of $1.3''\sim 23.8\: pc$. We find that the CO emission is clumpy and is distributed in two main structures: (i) the Horseshoe complex, located outside the H\rmnum{1} cloud, where gas is mostly excited by shocks and where no star formation is observed, and (ii) the Vertical filament, located at the edge of the H\rmnum{1} shell, which is a region of moderate star formation.
   We identified 140 molecular clouds using a clustering method applied to the CO data cube. %We used a Gaussian decomposition and a clustering method developed by \cite{MAMD_2017a} in order to extract the signal from the data and reconstruct a datacube of the CO emission. We then applied the \emph{clumpfind} algorithm on this datacube and extracted 140 giant molecular clouds.
A statistical study reveals that these clouds have very similar physical properties, such as size, velocity dispersion, and mass, as in the inner Milky Way. However, the range of radius available with the present ALMA observations does not enable us to investigate whether 
or not the clouds follow the Larson relation. The large virial parameter $\alpha_{vir}$ of the clouds suggests that gravity is not dominant and clouds are not gravitationally unstable.
   Finally, the total energy injection in the northern filaments of Centaurus A is of the same order as in the inner part of the Milky Way. The strong CO emission detected in the northern filaments is an indication that the energy injected by the jet acts positively in the formation of dense molecular gas. The relatively high virial parameter of the molecular clouds suggests that the injected kinetic energy is too strong for star formation to be efficient. This is particularly the case in the Horseshoe complex, where the virial parameter is the largest and where strong CO is detected with no associated star formation. This is the first evidence of AGN positive feedback in the sense of forming molecular gas through shocks, associated with low star formation efficiency due to turbulence injection by the interaction with the radio jet.}

\keywords{methods:data analysis - galaxies:individual:Centaurus A - galaxies:evolution - galaxies:interactions - galaxies:star formation - radio lines:galaxies}

\maketitle

%%%%%%%%%%%%%%%%%%%%%%%%%%%%%%%%%%%%%%%%%%%%%%%%%%%%%%%%%%%%%%%%%%%%%%%%%%%%%%%%%%%%%%%%%%%%%%%%%%%%%%%%%%%%

\section{Introduction}

   Active galactic nuclei (AGN) are assumed to play a major role in regulating star formation. The so-called AGN negative feedback is often invoked to explain the small number of massive galaxies compared to the predictions of the $\Lambda$-CDM model \citep{Bower_2006,Croton_2006,Harrison_2012,Dubois_2013, Werner_2014}. On the contrary, AGN with pronounced radio jets are prone to positive feedback \citep{Zinn_2013}. In particular, in high-redshift radio galaxies, optical emission was found to be aligned with the radio morphology. Such alignment is interpreted as the result of overpressured clouds in the intergalactic medium from the radio jet, triggering star formation in the direction of the radio jet \citep{McCarthy_1987,Begelman_1989,Rees_1989, deYoung_1989}.
Evidence of so-called jet-induced star formation was recently found at low \citep{vanBreugel_2004,Feain_2007,Inskip_2008,Elbaz_2009,Reines_2011,Combes_2015a} and high redshift \citep{Klamer_2004,Miley_2008,Emonts_2014}. Recent studies were also conducted with numerical simulations to study the effect of radio jets on star formation in the host galaxy or along the jet direction \citep{Wagner_2012,Gaibler_2012,Bieri_2016,Fragile_arXiv}.

   The widely studied nearby galaxy NGC 5128 (also known as Centaurus A) is the perfect target to study the radio jet-gas interaction. This galaxy hosts a massive disc of dust, gas, and young stars in its central regions \citep{Israel_1998} and is surrounded by faint arc-like stellar shells (at a radius of several kpc around the galaxy) in which H\rmnum{1} gas has been detected \citep{Schiminovich_1994}. Aligned with the radio jet, CO emission has been observed in the gaseous shells \citep{Charmandaris_2000}. In addition, \cite{Auld_2012} detected a large amount of dust ($\sim 10^5\: M_\odot$) around the northern shell region.

   Optically bright filaments are observed in the direction of the radio jet \citep{Blanco_1975,Graham_1981,Morganti_1991}. Galaxy Evolution Explorer (GALEX) data \citep{Auld_2012} and young stellar clusters \citep{Mould_2000,Rejkuba_2001} indicate that star formation occurs in these filaments. These so-called inner and outer filaments are located at a distance of $\sim 7.7\: kpc$ and $\sim 13.5\: kpc$ from the central galaxy, respectively. The inner and outer filaments show distinct kinematical components, i.e. a well-defined knotty filament and a more diffuse structure, as highlighted by optical excitation lines (VIMOS and MUSE; \citealt{Santoro_2015a, Santoro_2015b,Hamer_2015}). Recently \cite{Santoro_2016} identified a star-forming cloud in the MUSE data that contains several H\rmnum{2} regions. Some of these regions are currently forming stars, whereas star formation seems to have recently stopped in the others.

   In \cite{SalomeQ_2016b}, we mapped the outer filaments (hereafter the northern filaments) in $^{12}CO$(2-1) with APEX. The molecular gas was found to be very extended with a surprisingly CO bright region outside the H\rmnum{1} cloud. We found that star formation does not occur in this CO bright region. In the other part of the filaments, star formation appears to be inefficient compared to the Kennicutt-Schmidt law \citep{SalomeQ_2016a}. However, the jet-gas interaction seems to trigger the atomic-to-molecular gas phase transition \citep{SalomeQ_2016b}, suggesting that positive feedback is occuring in the filaments.

   The goal of this paper is to understand why the molecular gas in the northern filaments of Centaurus A does not follow the Kennicutt-Schmidt law, and whether there is a difference between the eastern CO bright region and the rest of the filaments. To do so, we decided to observe the CO emission seen with APEX at high resolution in order to resolved GMCs. In this paper, we present recent ALMA observations of the $^{12}CO$(1-0) line along the northern filaments of Centaurus A. The data and a method to extract clouds are presented in section \ref{sec:data}. In Section \ref{sec:Res}, we analyse the data and conduct a statistical analysis of the clouds. We discuss our results in section \ref{sec:discussion} and conclude in section \ref{sec:conclusion}.

\section{From CO emission to individual clouds}
\label{sec:data}

\begin{figure*}[h]
  \centering
  \includegraphics[height=9.5cm,trim=190 35 275 85,clip=true]{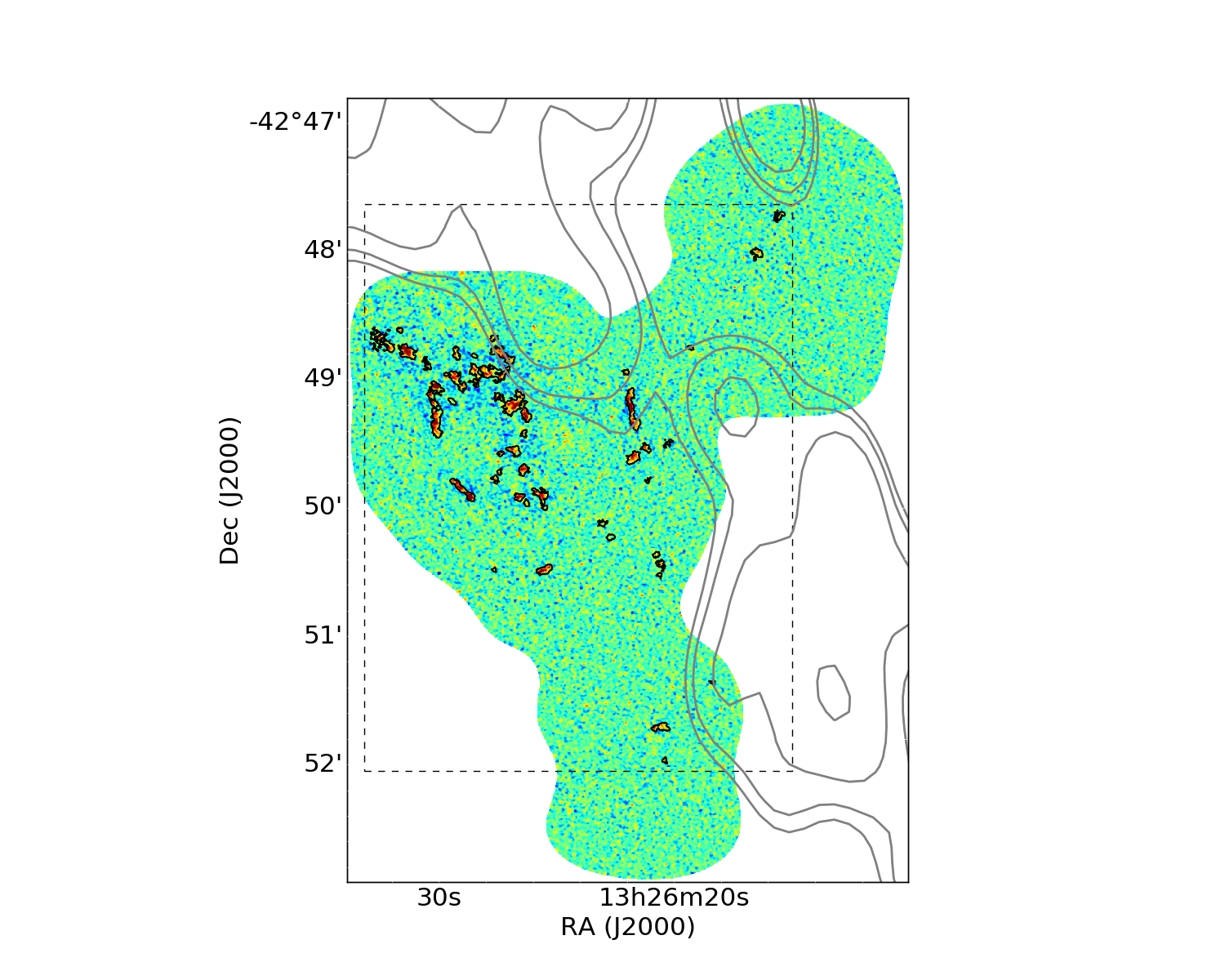}
  \hspace{3mm}
  \includegraphics[height=9.5cm,trim=275 80 195 80,clip=true]{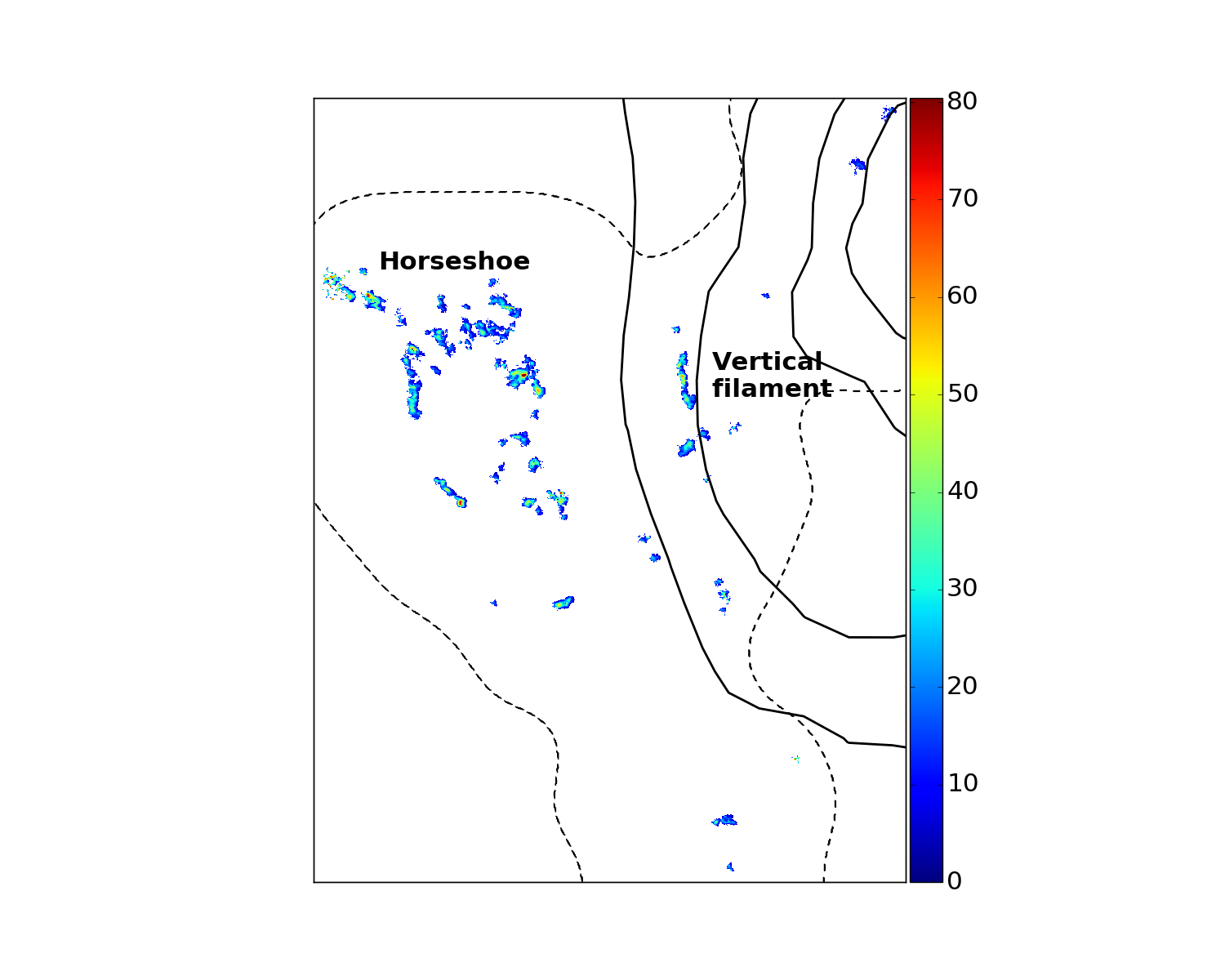} \\
  \vspace{3mm}
  \includegraphics[height=9.5cm,trim=275 80 185 85,clip=true]{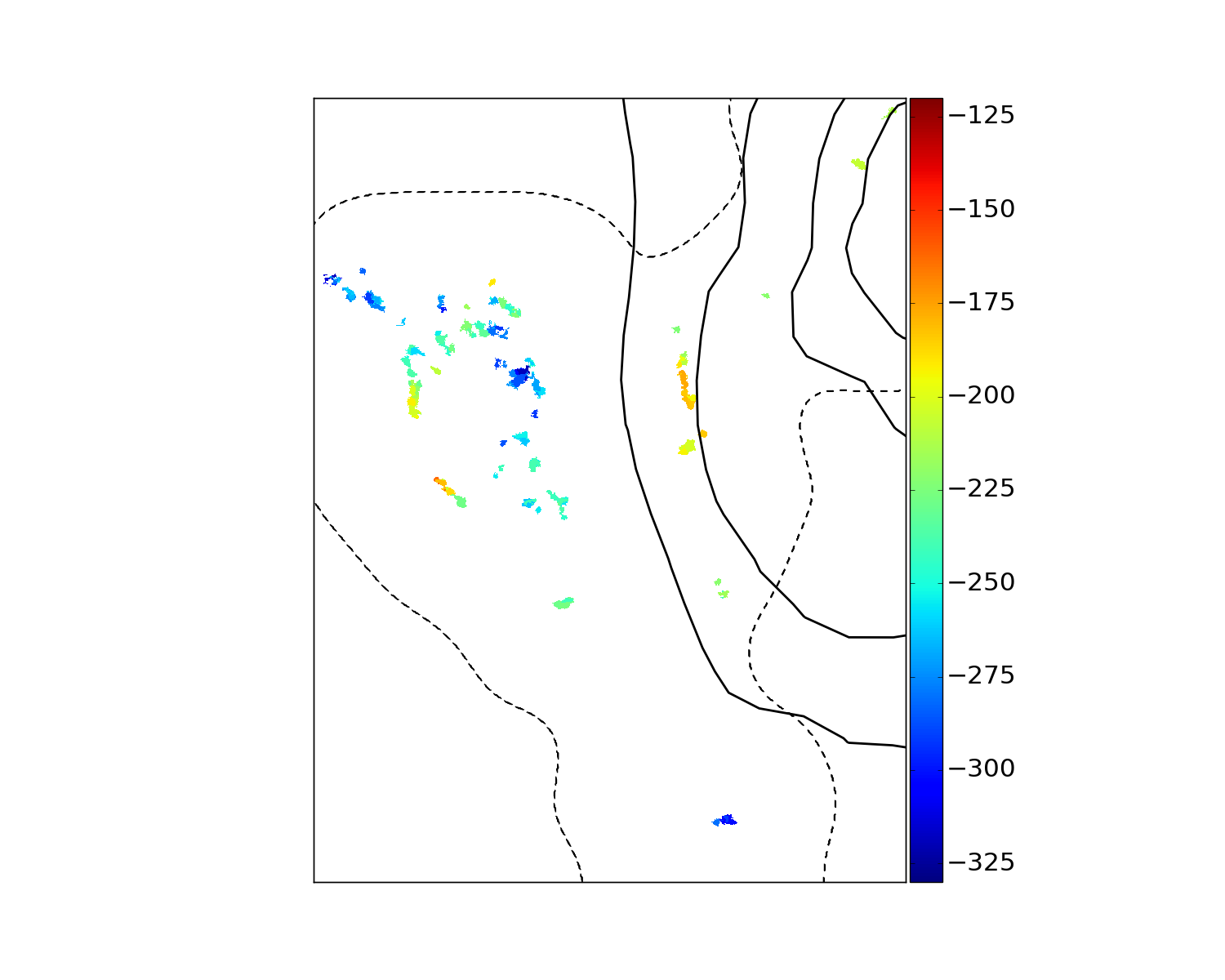}
  \hspace{3mm}
  \includegraphics[height=9.5cm,trim=275 80 195 85,clip=true]{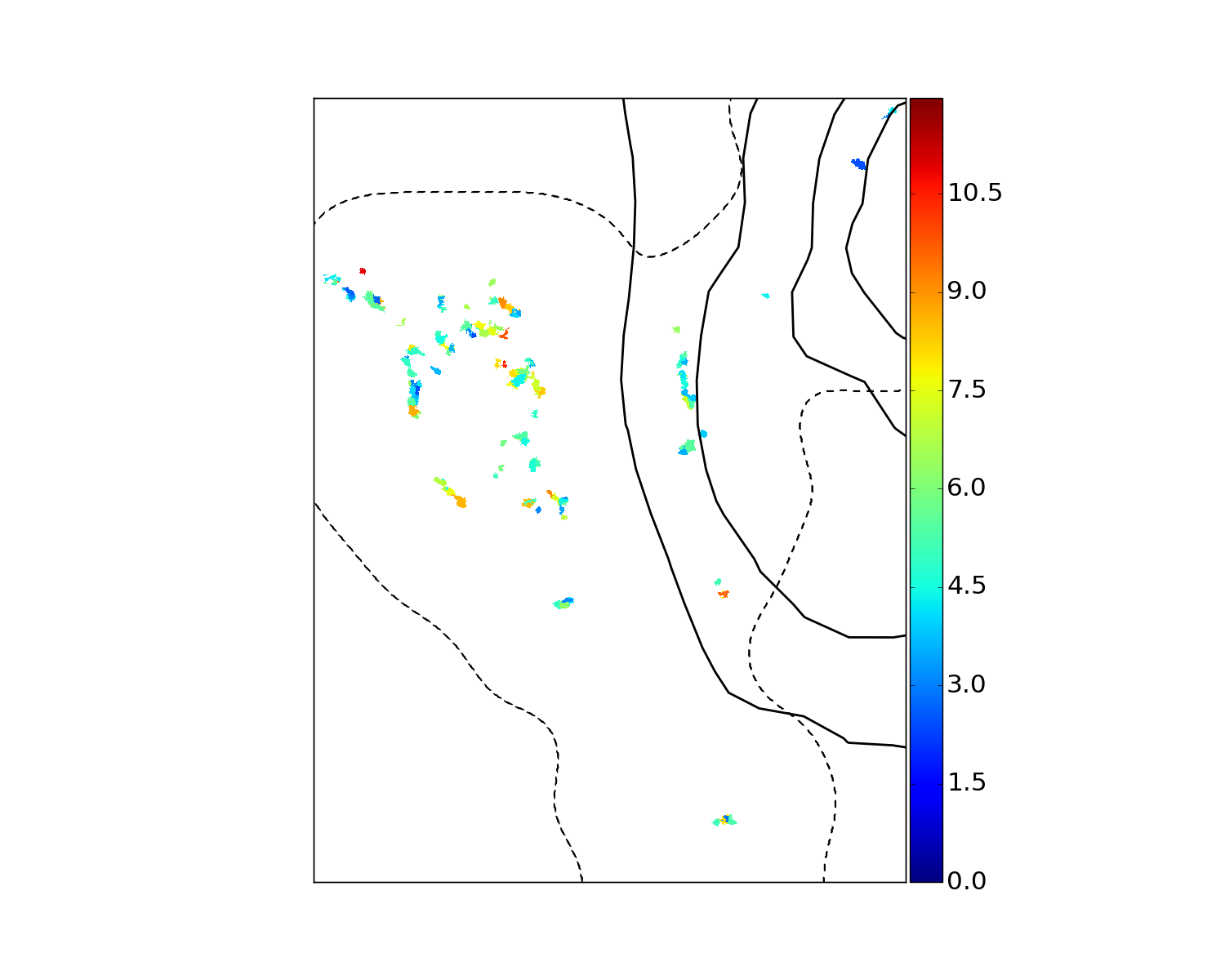}
  \caption{\label{extraction} \emph{Top left:} Map of the moment 0 of the ALMA data produced with CASA. We reported, in black contours, the structures extracted by the method presented in section \ref{sec:extraction}.  We indicate the location of the radio jet in grey contours. \emph{Top right:} Integrated CO intensity map in $mJy.km.s^{-1}$ of the molecular clouds seen in CO(1-0) with ALMA. \emph{Bottom:} Maps of the central velocity relative to Centaurus A (\emph{left}) and velocity dispersion (\emph{right}) in $km.s^{-1}$ of the clouds. The area of the three maps corresponds to the dashed box in the moment 0 map. The full line and dashed contours represent the H\rmnum{1} emission \citep{Schiminovich_1994} and the region observed with ALMA, respectively. The CO emission is clumpy and covers only a small fraction of the region mapped with ALMA.}
\end{figure*}

   \subsection{ALMA data}
   \label{sec:Obs}

   We mapped the $^{12}CO$(1-0) emission in the northern filaments of Centaurus A with the 12 m array of the Atacama Large Millimeter/submillimeter Array (ALMA) during Cycle 3 using Band 3 receivers (project ADS/JAO.ALMA$\#$2015.1.01019.S). The map covers a region of 6.1$'\times $4.3$'$ and consists in a mosaic of 34 pointings, each with an integration time between 140 and 430s. The baselines ranged from 15 m to 704 m, providing a resolution of $1.30''\times 0.99''$ ($PA=81.5$\degree). At the distance of Centaurus A (3.8 Mpc, $1''\sim 18.3\: pc$; \citealt{Harris_2010}), this corresponds to a beam size of $23.8\times 18.1\: pc$. The largest angular scale recovered by the interferometer is $14''$ , which corresponds to about 260 pc at the distance of Centaurus A.

   The data were calibrated using the Common Astronomy Software Applications (CASA) and the supplied script. Owing to the primary beam correction, the noise level of the map is not uniform and higher on the edge of map. The histogram of the noise level peaks at 6.5 mJy/beam at a spectral resolution of $1.47\: km.s^{-1}$.

   \subsection{Cloud identification method}
   \label{sec:extraction}

   To identify clouds in the ALMA data, we first used a Gaussian decomposition method developed by \cite{MAMD_2017a}. This algorithm allows recovery of the signal, even at low signal-to-noise ratio. Clustering is then made with a threshold descent similar to the \emph{clumpfind} algorithm \citep{Clumpfind}, but applied on a cube of the integrated flux (see \citealt{MAMD_2017a} for the details).

   We kept the structures with a size larger or equal to the beam size. We also discarded those with central velocities outside the range $-350\leq v_{cent}\leq -100\: km.s^{-1}$ (where CO(2-1) has been detected; \citealt{SalomeQ_2016b}) or velocity dispersion higher than $\sigma_v\sim 50\: km.s^{-1}$.
The clustering method we used enables us to reconstruct a data cube of the modelled signal. We then applied the \emph{clumpfind} algorithm \citep{Clumpfind} on this cube with five equally spaced threshold values from 6.5 to 58.5 mJy/beam. After deleting spurious clumps smaller than the beam, we obtained a catalogue of 140 GMCs. The map of the integrated CO intensity (Fig. \ref{extraction}) shows small bright spots that are likely associated with barely resolved giant molecular clouds (GMCs).  Figure \ref{extraction} also shows the maps of the central velocity and velocity dispersion of the GMCs.

   \subsection{Effect of large-scale filtering}

   In \cite{SalomeQ_2016b}, we observed the $^{12}CO$(2-1) line with APEX along the northern filaments of Centaurus A. We determined that the total molecular gas mass is $(9.8\pm 0.6)\times 10^7\: M_\odot$. With ALMA, the mass derived from the $^{12}CO$(1-0) emission is much smaller $2.5\times 10^7\: M_\odot$; i.e. $1.73\times 10^7\: M_\odot$ when we exclude three structures that lie outside the region previously observed with APEX.
In particular, in the CO bright region discovered by \cite{SalomeQ_2016b}, the total mass recovered by ALMA is $1.24\times 10^7\: M_\odot$, a factor of 5 smaller than the mass derived from the CO(2-1) from APEX $(6.3\pm 0.2)\times 10^7\: M_\odot$.

   Such a difference can be partly explained by short-spacing filtering. The northern filaments have also been mapped with the Atacama Compact Array (ACA) during Cycle 3. These data will be presented in a future paper, however we already determined that the molecular gas mass is a factor of 4 higher than that derived with the data from the ALMA 12 m array alone.
The difference of mass between ALMA and APEX may also result from the CO(2-1)/CO(1-0) ratio used to derive the mass with APEX (0.55; following \citealt{Charmandaris_2000}). Taking into account the factor of 4 due to the short-spacing filtering, we estimate that the CO(2-1)/CO(1-0) ratio is about 0.7-0.8. This will be the subject of a third forthcoming paper, where we will compare the CO(1-0) emission from ALMA with high-J CO transitions observed with APEX in the CO bright region, and discuss the excitation of gas in this region.

\section{Results}
\label{sec:Res}

   \subsection{Spatial distribution of molecular clouds}

\begin{figure*}[h!]
  \centering
  \includegraphics[width=0.45\linewidth,trim=215 35 275 85,clip=true]{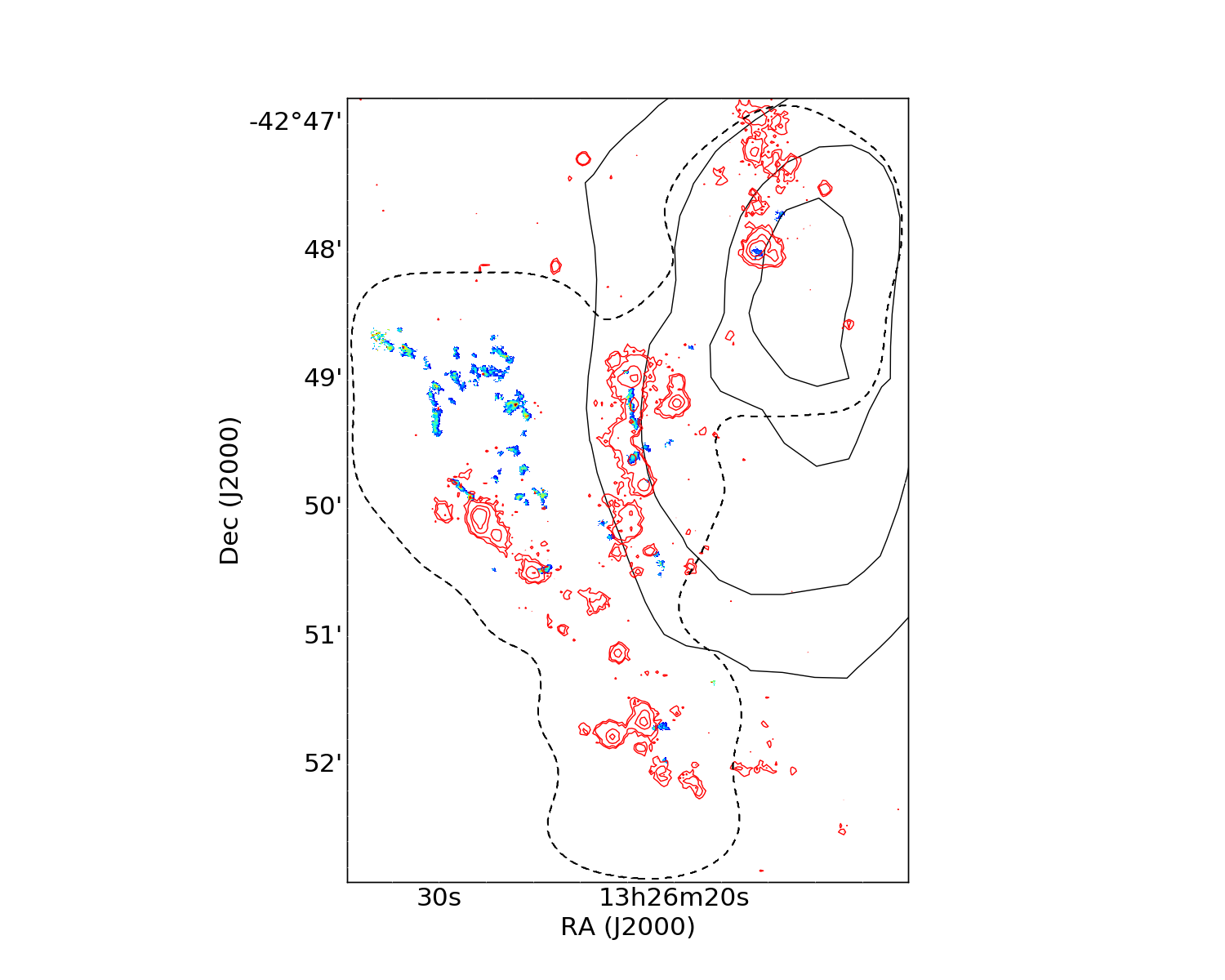}
  \includegraphics[width=0.45\linewidth,trim=215 35 275 85,clip=true]{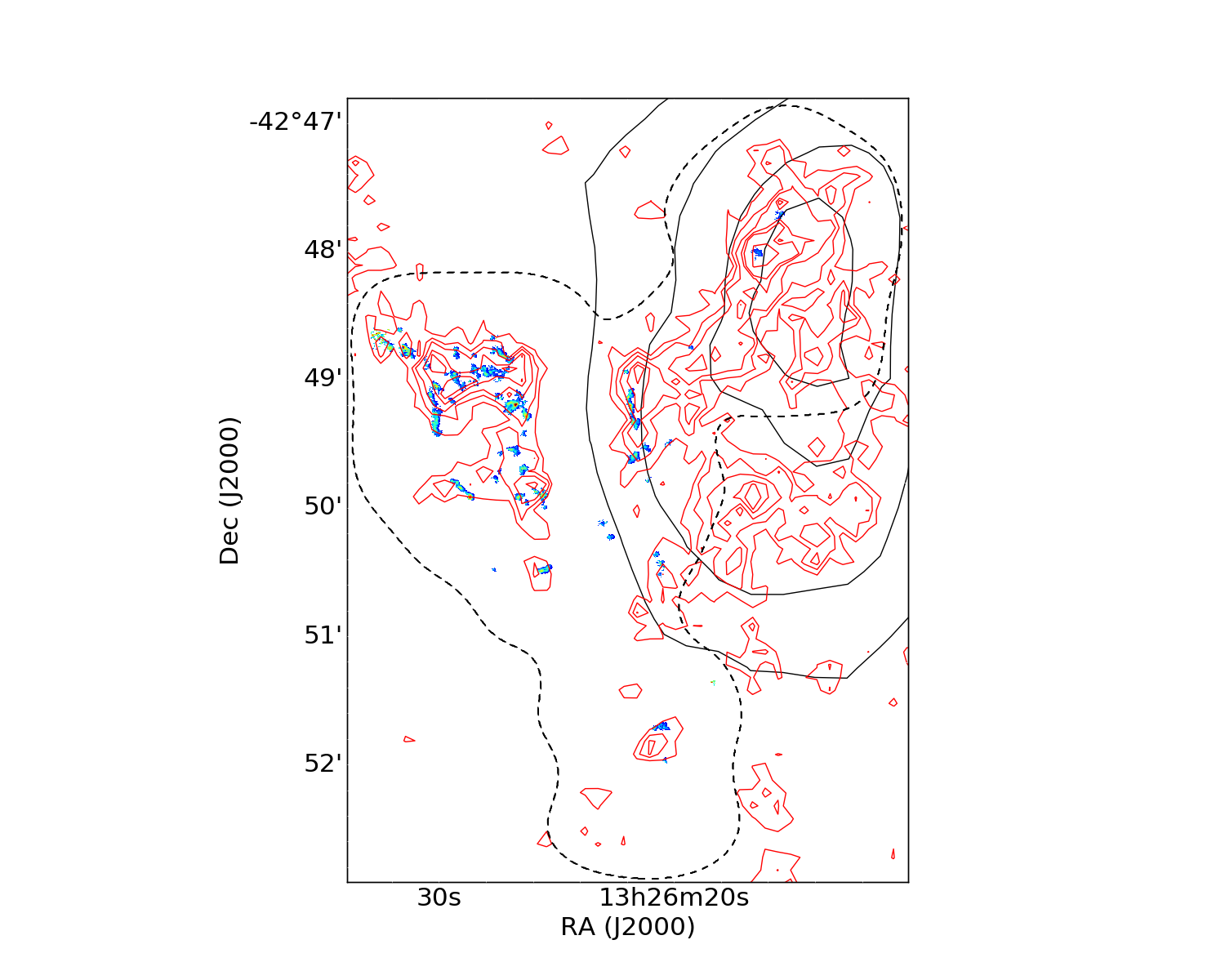} \\
  \vspace{3mm}
  \includegraphics[width=0.45\linewidth,trim=215 35 275 85,clip=true]{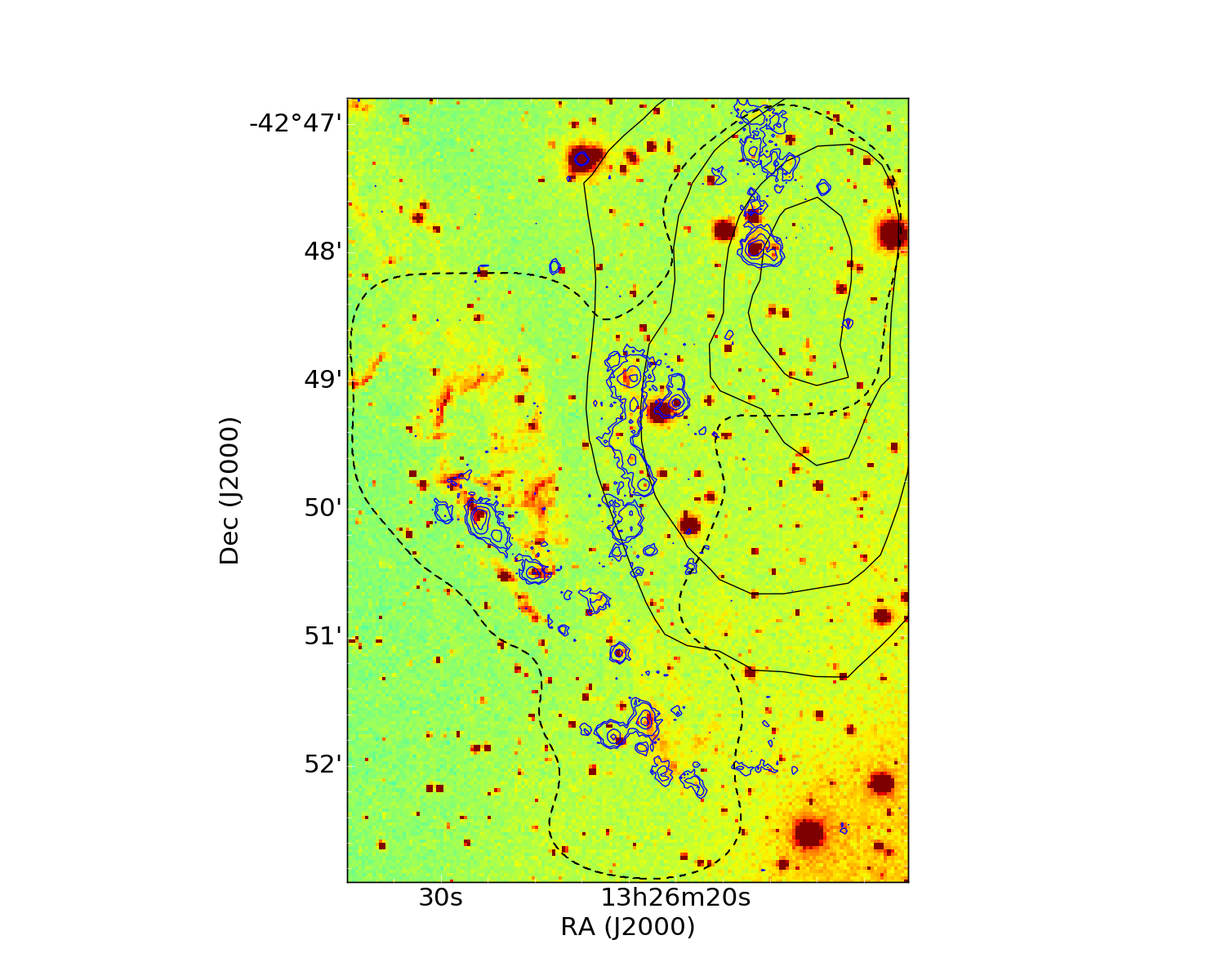}
  \includegraphics[width=0.45\linewidth,trim=215 35 275 85,clip=true]{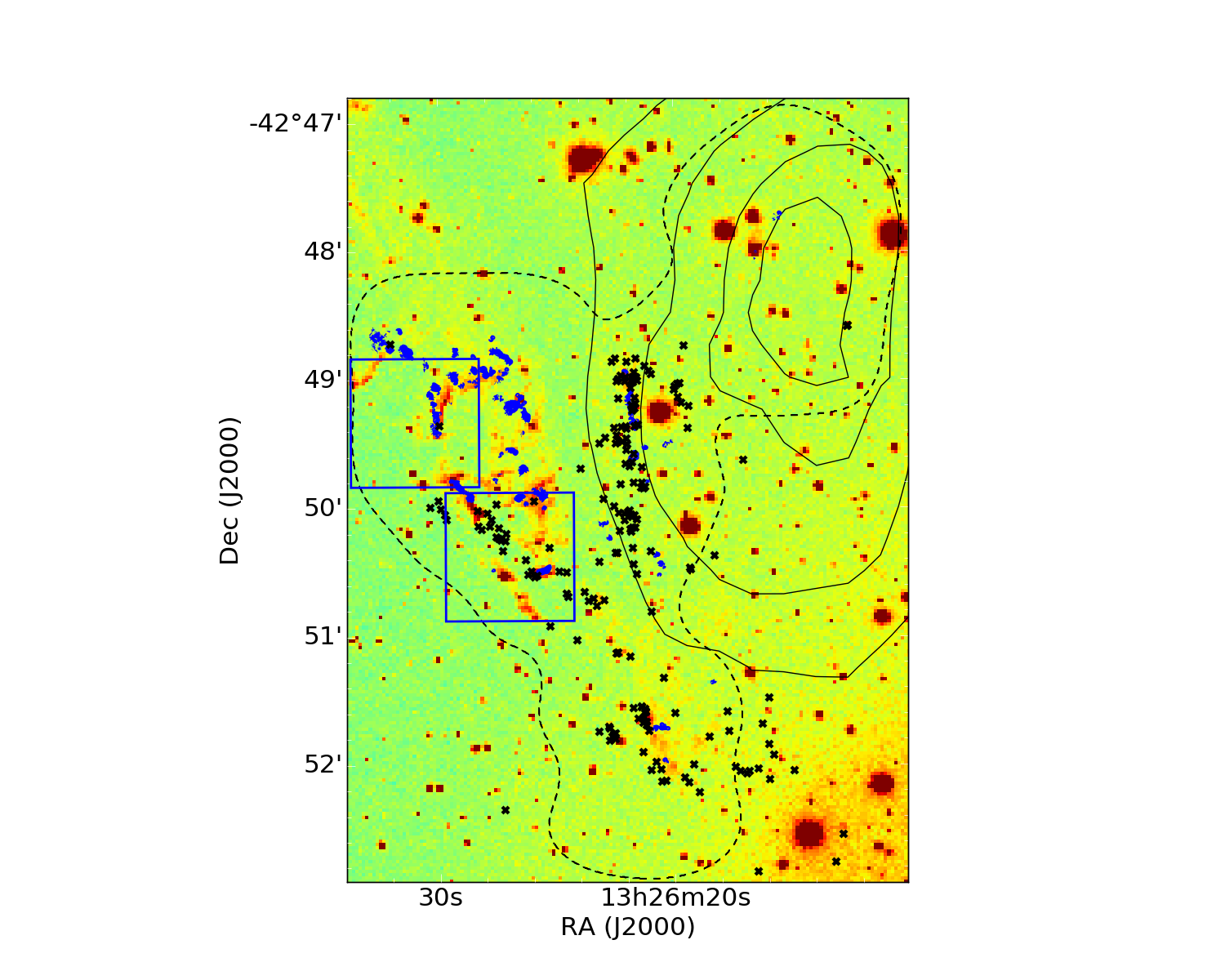}
  \caption{\label{multi_wl} Distribution of the ALMA CO(1-0) emission in relation to star formation regions. \emph{Top:} CO intensity map with contours of the GALEX FUV (\emph{left} - \cite{Neff_2015b}) and Herschel FIR (\emph{right}) emission. \emph{Bottom:} The colour map represents the $H\alpha$ emission seen with CTIO. In all four images, the black contours represent the H\rmnum{1} emission and the black dashed line corresponds to the area observed with ALMA. The crosses on the bottom right image indicate the position of the young star clusters from \cite{Rejkuba_2001}. \\
The Horseshoe complex, which corresponds to the bright CO emission from \cite{SalomeQ_2016b}, is not associated with young stellar clusters. In contrast, the Vertical filament (at the edge of the H\rmnum{1} shell) is associated with $H\alpha$, FUV emission, and young stellar clusters \citep{Rejkuba_2001}, and is likely forming stars. It is not established whether the northern FUV emission is associated with young stars as it was not included in \cite{Rejkuba_2001}.}
\end{figure*}

   Figure \ref{extraction} shows the maps of the CO(1-0) intensity, central velocity, and velocity dispersion of the clouds extracted with the method presented in section \ref{sec:extraction}). The distribution shows that most of the gas is distributed in low filling factor structures over the whole region, at the present noise level ($rms\sim 113\: mJy.km.s^{-1}$), as suggested by the three distinct unresolved and dynamically separated clumps previously found in archival ALMA CO(2-1) data \citep{SalomeQ_2016a}. The intensity map reveals the clumpy structure of molecular gas in the northern filaments of Centaurus A from which we could identify different complexes.
\medskip

   As expected from previous APEX data \citep{SalomeQ_2016b}, most of the CO(1-0) emission comes from the eastern part of the filaments (almost 77\% of the mass). In this region, the velocity map (Fig. \ref{extraction} - bottom left) shows the possible existence of coherent filamentary structures that present a horseshoe-like shape, along which molecular clouds are distributed. The clouds in the Horseshoe complex have higher velocity dispersions (Fig. \ref{extraction} - bottom right) than in the other molecular clouds of the northern filaments.

   To put the CO emission in context, Fig. \ref{multi_wl} presents the spatial relationship with other tracers, namely far-UV (FUV), dust emission, $H\alpha$, H\rmnum{1,} and young star clusters. In the eastern region, the structure observed in CO follows the dust emission observed with Herschel with a similar morphology. The Horseshoe complex is also associated with a similar structure seen in $H\alpha$ emission. In contrast, there is no FUV emission associated with the Horseshoe complex or young stellar clusters.

\begin{figure*}[h!]
  \centering
  \includegraphics[page=1,width=0.49\linewidth,trim=15 0 45 30,clip=true]{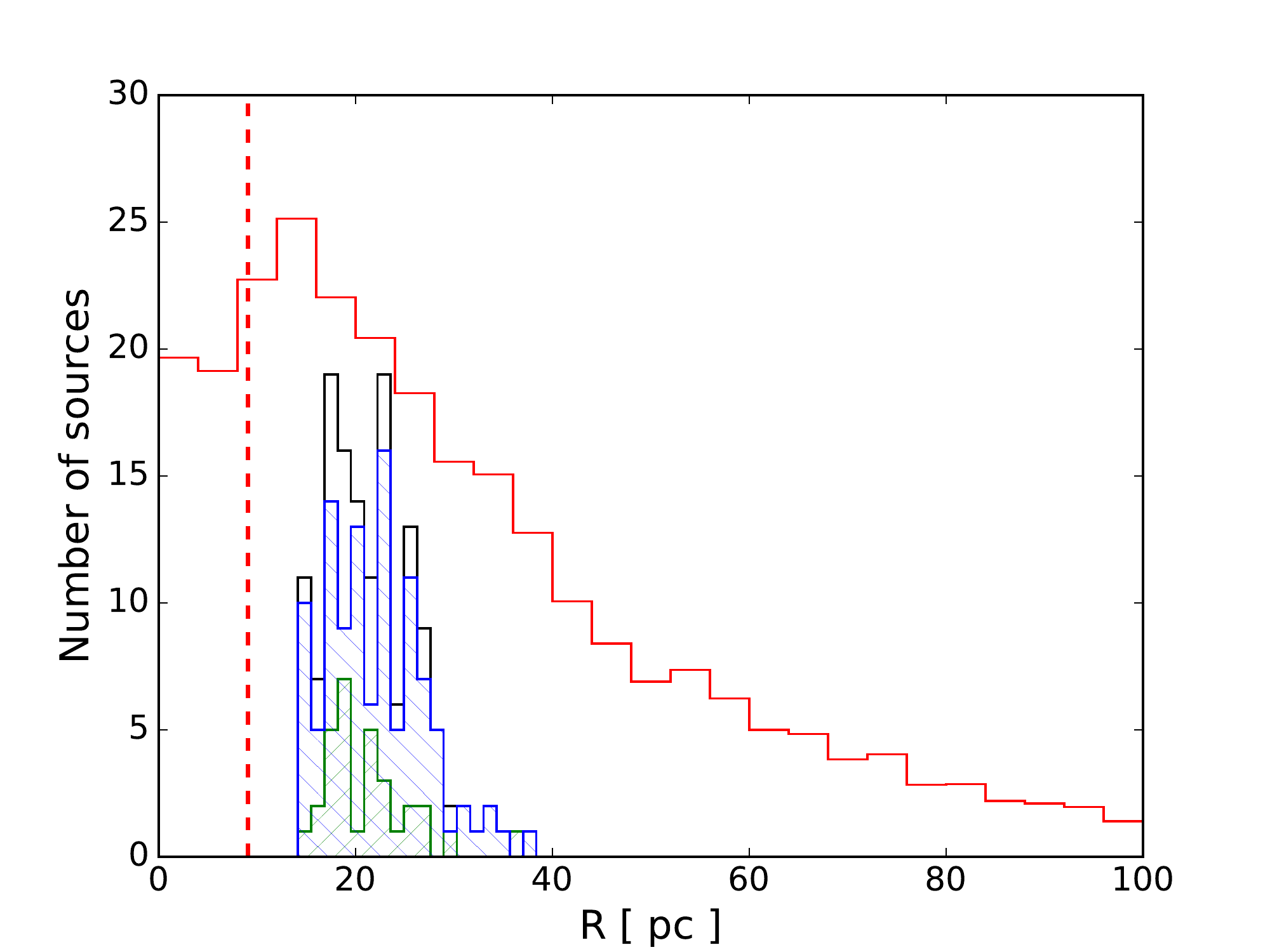}
  \includegraphics[page=2,width=0.49\linewidth,trim=15 0 45 30,clip=true]{Statistics_clumps.pdf} \\
  \includegraphics[page=4,width=0.49\linewidth,trim=15 0 45 30,clip=true]{Statistics_clumps.pdf}
  \includegraphics[page=5,width=0.49\linewidth,trim=15 0 45 30,clip=true]{Statistics_clumps.pdf} \\
  \includegraphics[page=6,width=0.49\linewidth,trim=15 0 45 30,clip=true]{Statistics_clumps.pdf}
  \includegraphics[page=7,width=0.49\linewidth,trim=15 0 45 30,clip=true]{Statistics_clumps.pdf}
  \caption{\label{Hist} Histogram of the properties of the giant molecular clouds. \emph{Top:} Characteristic radius and velocity dispersion. \emph{Middle:} Molecular gas mass and surface density. The red dashed line in the left plot is the best fitting power law for masses $M\geq 9\times 10^4\: M_\odot$. \emph{Bottom:} Volume density and virial parameter. The vertical red dashed lines correspond to the lower limits due to resolution or sensitivity. In the six plots, we added the histograms of GMCs of the Milky Way \citep{MAMD_2017a}.
The clouds were separated in two groups: (i) the Horseshoe complex in blue, where gas is excited by shocks; and (ii) the star-forming clouds in green, associated with $H\alpha$, FUV emission, and young stellar clusters.}
\end{figure*}

   While CO emission covers all the dust emission in the eastern region, it only covers a small fraction of dust emission in the H\rmnum{1} cloud. Most of the CO emission in the H\rmnum{1} cloud is distributed in a vertical filament located at the edge of the H\rmnum{1} cloud. This filament is likely a single coherent structure as it does not show significant difference in the central velocity (Fig. \ref{extraction}). This filament also does not seem to be highly turbulent with velocity dispersions lower than $6-7\: km.s^{-1}$.
The Vertical filament follows a filament seen in $H\alpha$ with CTIO (Fig. \ref{multi_wl}), and is also aligned with FUV emission that is likely produced by young stellar clusters found by \cite{Rejkuba_2001}. This tends to indicate that the Vertical filament is a region of star formation, contrary to the Horseshoe complex where the $H\alpha$ emission is mostly excited by shocks \citep{SalomeQ_2016b}.

   In addition to the Vertical filament, the H\rmnum{1} cloud also contains a few isolated structures seen in CO. In particular, the peak of H\rmnum{1} emission is associated with only two small CO structures. Interestingly, all the isolated CO structures within the H\rmnum{1} cloud may be star-forming regions as they are associated with FUV emission and young stellar clusters. This also seems to be the case for the CO emission located in the south that is associated with a bright spot of both FUV and dust emission.

   The molecular clouds observed with ALMA seem to present two star formation regimes. Molecular gas in the Horseshoe complex forms stars very inefficiently and the $H\alpha$ emission is excited by shocks, whereas in the Vertical filament and isolated clouds the CO emission is associated with recent star formation.

   \subsection{Giant molecular clouds}
   \label{sec:prop}

   In this section we discuss the physical properties of the clouds, looking for differences between the Horseshoe complex and star-forming regions. In Fig. \ref{Hist} and Table \ref{table:stats}, we report the statistics of the radius, velocity dispersion, mass, surface density, and density.
\medskip

\noindent \textit{Size} - The angular radius is defined as the emission-weighted radius. We used the implementation made by \cite{MAMD_2017a} that is based on the inertia matrix
\begin{equation}
  \mathnormal{\psi=
  \begin{bmatrix}
    \sigma_\alpha^2         & \sigma_{\alpha\delta}^2 \\
    \sigma_{\alpha\delta}^2 & \sigma_\delta^2
  \end{bmatrix}
  }
,\end{equation}
where
\begin{eqnarray}
  \mathnormal{\sigma_\alpha^2}  &=&\mathnormal{\frac{\sum_{pix} S^i_{CO} (\alpha-\alpha_i)^2}{\sum_{pix} S^i_{CO}}} \\
  \mathnormal{\sigma_\delta^2}  &=&\mathnormal{\frac{\sum_{pix} S^i_{CO} (\delta-\delta_i)^2}{\sum_{pix} S^i_{CO}}} \\
  \mathnormal{\sigma_{\alpha\delta}^2} &=&\mathnormal{\frac{\sum_{pix} S^i_{CO} (\alpha-\alpha_i)(\delta-\delta_i)}{\sum_{pix} S^i_{CO}}}
\end{eqnarray} 
with $\alpha$,$\delta$ the central coordinates of the cloud.
The size is given by the eigenvalues of $\mathnormal{\psi}$, where the largest and smallest half-axis of the cloud $R_{min}$ and $R_{max}$ are the maximum and minimum eigenvalues. We assume that the GMCs are prolate and adopt the following definition of the angular radius:
\begin{eqnarray}
  R_{ang}=(R_{max} R_{min} R_{min})^{1/3}.
\end{eqnarray}
Finally, the physical radius is given by $R=D_L\tan(R_{ang})$, where $D_L$ is the luminosity distance. The minimum radius found is limited by the resolution ($R\sim 14.1\: pc$, about 1.4 times the resolution). The distribution of radius is rather narrow; 54\% of the GMCs are larger than twice the resolution and smaller than three times the resolution (Fig. \ref{Hist} - top left). The larger value of R is 38.4 pc, about four times the resolution. This narrow range is due to the way \emph{clumpfind} identifies structures. Because of lack of larger scale structures in the data (due to interferometry filtering), \emph{clumpfind} tends to split potentially larger coherent structures in a collection of individual clumps, missing associations. Adding short-spacings will enable us to determine more robustly the spatial-scale distribution of molecular gas.
\medskip

\noindent \textit{Velocity dispersion} - We defined the central velocity of the clouds as the emission-weighted mean velocity $\avg{v}$ and the velocity dispersion $\sigma_v$ as follows:
\begin{eqnarray}
  \avg{v}  &=&\mathnormal{\frac{\sum_v v\, S_{CO}(v)\, dv}{\sum_v S_{CO}(v)}} \\
  \sigma_v &=&\sqrt{\mathnormal{\frac{\sum_v v^2\, S_{CO}(v)\, dv}{\sum_v S_{CO}(v)}-\avg{v}^2}}
,\end{eqnarray} 
where $\mathnormal{S_{CO}(v)}$ is the average CO spectrum of a single cloud, constructed by adding all the voxels of the data cube identified by \emph{clumpfind}, and $\mathnormal{dv}$ is the channel width. The GMCs have velocity dispersions lower than $11\: km.s^{-1}$, with an average value of $5.4\pm 1.4\: km.s^{-1}$.
\medskip

\noindent \textit{Mass} - The CO luminosity was derived using the relation of \cite{Solomon_1987},
\begin{equation}
  L'_{CO}=3.25\times 10^7\, S_{CO}\Delta v\, D_L^2\, \nu_{obs}^{-2} (1+z)^{-3}
,\end{equation}
where $\Delta v$ is the FWHM and $S_{CO}\Delta v=\sum_i S_{CO}\Delta v_i$ is the total integrated emission of the cloud.
The mass is then estimated by applying a CO-to-$H_2$ conversion factor of $\alpha_{CO}=4.3\: M_\odot.(K.km.s^{-1}.pc^2)^{-1}$ \citep{Bolatto_2013}. The total mass of the GMCs is $\sim 1.6\times 10^7\: M_\odot$. This represents about 63\% of the mass extracted from the CO(1-0) ALMA data by the Gaussian decomposition and clustering methods.
The mass distribution of the clouds is shown in Fig. \ref{Hist} (middle left). The mass ranges from $7.8\times 10^3$ to $6.1\times 10^5\: M_\odot$ with a mean value of $1.1\times 10^5\: M_\odot$. We fitted the high mass part of the distribution using a power-law $dN/dlog(M)\propto M^{-\alpha}$ assuming a $1/\sqrt{N}$ uncertainty for each data point. For $M\geq 9.3\times 10^4\: M_\odot$ we obtain $\alpha=0.75\pm 0.18$.
This value is consistent with that found for giant molecular clouds in general ($\alpha\sim 0.8$ for $M>10^4\: M_\odot$, \citealt{Solomon_1987,Kramer_1998, Heyer_2001,Marshall_2009}).
\medskip

\begin{table*}[h]
  \centering
  \small
  \caption{\label{table:stats} Statistics (mean and mean absolute deviation) of the giant molecular clouds properties. For comparison, we list the properties of clouds in M 33, M 51, the Large Magellanic Cloud, and the Milky Way \citep{Hughes_2013,MAMD_2017a}.}
  \begin{tabular}{lccccccc}
    \hline \hline
    Quantity                           &   Horseshoe    &   SF clouds    &     M33      &     LMC      &     M51      &  MW  \\ \hline
    Radius (pc)                        & $22.0\pm 4.1$  & $21.2\pm 3.3$  &  $51\pm 13$  &  $16\pm 5$   &  $48\pm 14$  & 31.5 \\
    $\sigma_v$ ($km.s^{-1}$            &  $5.6\pm 1.6$  &  $4.9\pm 1.2$  & $3.8\pm 0.7$ & $1.6\pm 0.4$ & $6.4\pm 1.8$ &  4.0 \\
    Mass ($10^5\:M_\odot$)             &  $1.1\pm 0.8$  &  $1.2\pm 0.8$  &      -       &      -       &      -       &  1.5 \\
    $\Sigma_{H_2}$ ($M_\odot.pc^{-2}$) & $45.1\pm 18.2$ & $51.3\pm 21.0$ &  $46\pm 20$  &  $21\pm 9$   & $180\pm 82$  & 28.6 \\
    Density ($cm^{-3}$)                & $37.5\pm 18.4$ & $45.0\pm 22.1$ &      -       &      -       &      -       & 24.1 \\
    $\alpha_{vir}$                     & $12.0\pm 7.0$  &  $8.0\pm 4.4$  & $2.1\pm 0.9$ & $1.6\pm 0.3$ & $1.7\pm 0.4$ &  -   \\ \hline
  \end{tabular}
\end{table*}

\noindent \textit{Surface density} - The angular area of a cloud is defined as $A=N_{pix}d\Omega$, where $N_{pix}$ is the number of pixels on the sky and $d\Omega$ is the solid angle of a single pixel. Dividing the mass by this area in squared parsecs, we obtain the surface density $\Sigma_{H_2}$.
The histogram of the GMCs surface density is rather narrow with an average value of $46.5\pm 19.1\: M_\odot.pc^{-2}$. In particular, the histogram does not show low surface densities comparable to the outer Milky Way. This is likely due to the sensitivity of the ALMA observations. At the present noise level, the lower limit detectable at $3\sigma$ is $I_{CO}>1.75\: K.km.s^{-1}$, which corresponds to a surface density $\Sigma_{H_2}>7.64\: M_\odot.pc^{-2}$ (the vertical red dashed line on the middle right panel of Fig. \ref{Hist}).

\begin{figure}[h]
  \centering
  \includegraphics[page=3,width=\linewidth,trim=20 5 45 35,clip=true]{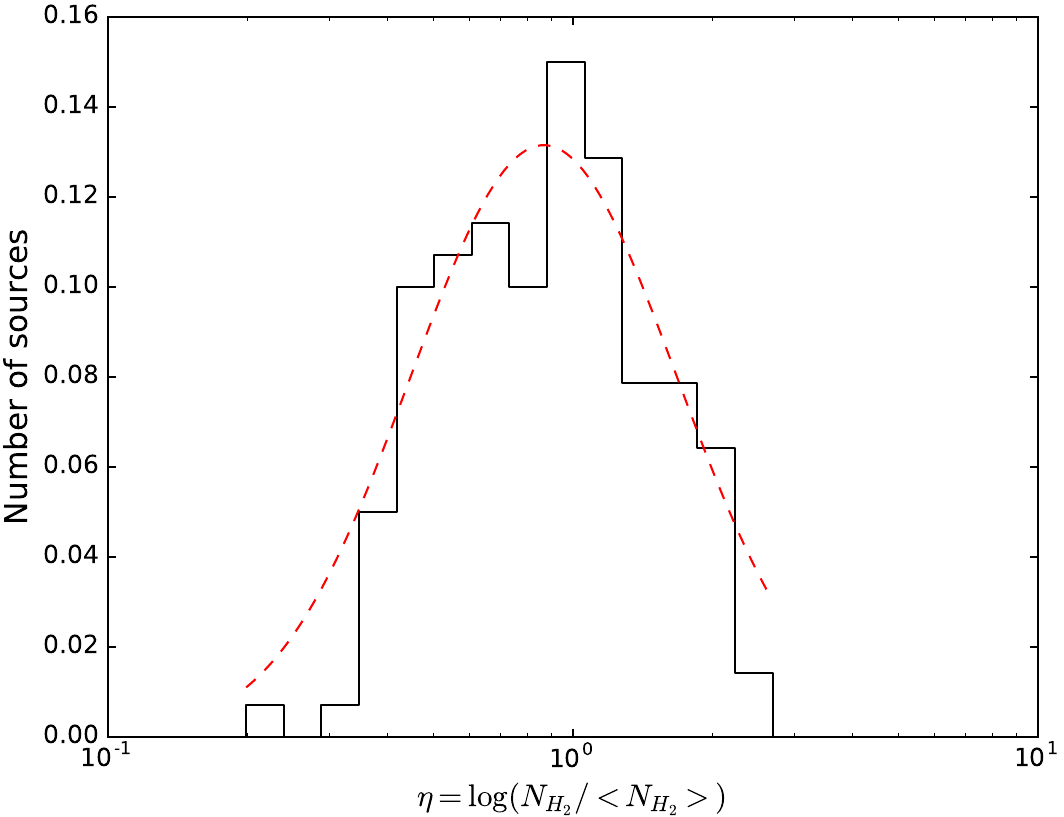}
  \caption{\label{PDF} Probability distribution function of the $H_2$ column density of the clouds. The x-axis shows the normalised column density. The red dashed line shows the best fitting log-normal ($\sigma=0.66\pm 0.08$, $\mu=0.29\pm 0.11$).}
\end{figure}

   The probability distribution function of the surface density shows a log-normal shape (Fig. \ref{PDF}). The best fitting model has a location parameter $\mu=0.29\pm 0.11$ and a scale parameter $\sigma=0.66\pm 0.08$ (see Fig. \ref{PDF}). Using numerical simulations of compressive supersonic flows, \cite{Kritsuk_2007,Kritsuk_2011} showed that the surface density histogram is found to be log-normal in the absence of gravity. When including gravity in the simulations, the histogram at high density deviates from log-normal in the form of a power-law tail. In the northern filaments of Centaurus A, the PDF of the GMCs surface density does not show a power-law behaviour at high surface density. This suggests that the contribution of gravitation is small.
\medskip

\begin{figure*}[h]
  \centering
  \includegraphics[page=1,width=0.4\linewidth,trim=25 5 45 35,clip=true]{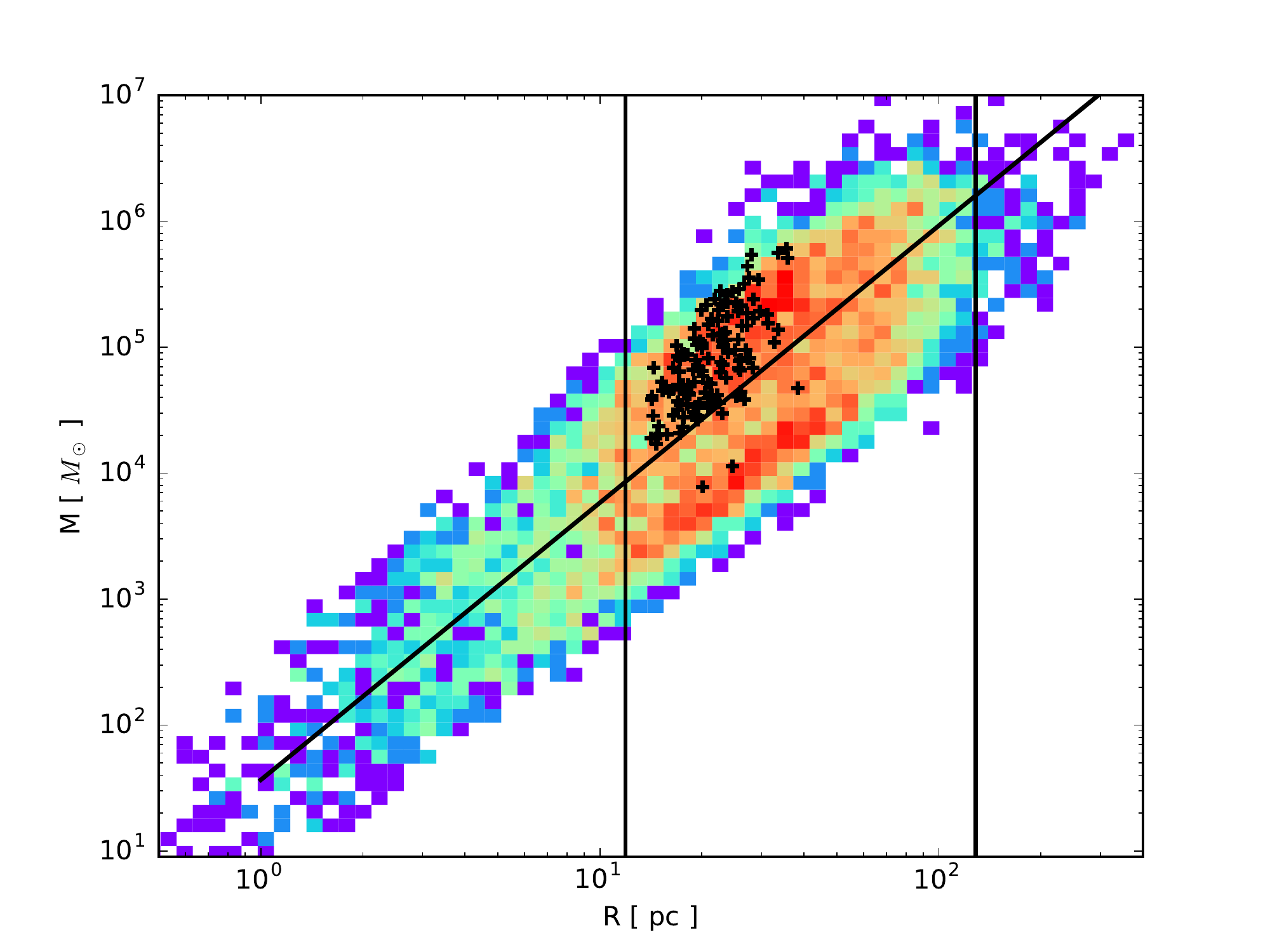}
  \includegraphics[page=3,width=0.4\linewidth,trim=25 5 45 35,clip=true]{Larson_relations.pdf}
  \includegraphics[page=4,width=0.4\linewidth,trim=25 5 45 35,clip=true]{Larson_relations.pdf}
  \includegraphics[page=5,width=0.4\linewidth,trim=25 5 45 35,clip=true]{Larson_relations.pdf}
  \caption{\label{Larson} Plot of the Larson's relations for the giant molecular clouds (black points). The top row shows the $M-R$ (\emph{left}) and $\sigma-R$ (\emph{right}) relations. The bottom row shows the $\sigma-M$ (\emph{left}) and $\sigma-\Sigma R$ (\emph{right}) relations. We also plot the 2D histograms and the best fitting relations (colour scale and black line) for the molecular clouds in the Milky Way \citep{MAMD_2017a}. The molecular clouds in the northern filaments of Centaurus A are consistent with those of the Milky Way.
We also indicate the limits on the radius and velocity dispersion that can be reached owing to resolution or filtering effects ($18.1\leq R\leq260\: pc$, $\sigma_{v}>1.25\: km.s^{-1}$; horizontal and vertical black lines).}
\end{figure*}

\noindent \textit{Volume density} - The gas density of each cloud is defined by
\begin{equation}
  n_H=\frac{3M}{4\pi R^3}\frac{1}{(\mu+f_{H_2})m_H} \label{eq:density}
,\end{equation}
where $\mu=2.4$ to take heavy elements into account and $f_{H_2}$ is the molecular gas fraction. As previously found by \cite{SalomeQ_2016b}, the filaments are mostly molecular thus, we assume $f_{H_2}=1$.
The GMCs cover a large range of densities from 3.0 to $98.8\: cm^{-3}$ with an average density of $39.2\pm 19.3\: cm^{-3}$ (see Fig. \ref{Hist} - bottom left).

   \subsection{Larson's relations}

\noindent \textit{Mass-size relation} - The top left panel of Fig. \ref{Larson} shows the relation between the mass and the size of the CO clouds. We added the 2D histograms of the molecular clouds in the Milky Way \citep{MAMD_2017a}. The giant molecular clouds of the northern filaments of Centaurus A are consistent with the dispersion for molecular clouds in the inner Milky Way.
We tried a bisector linear regression Y versus X and X versus Y to estimate the variation of mass with the radius. The index of the power law is larger than that found for the Milky Way \citep{MAMD_2017a}. However, the dispersion of the points in the $M-R$ plot is of the same order as the range of radius available with the present ALMA observations (less than one order of magnitude), therefore fitting a linear relation is not really significant yet.
\medskip

\noindent \textit{Velocity-size relation} - The $\sigma_v-R$ relation is shown in the top right panel of Fig. \ref{Larson}. For this plot, the data points also agree with the 2D histogram of the Milky Way distribution. However, there is no correlation between $\sigma_v$ and R for the molecular clouds of the northern filaments. Again, this is likely due to the small coverage in the spatial scale of the present observations.
In addition to $\sigma_v-R$, we explored various relations between $\sigma_v$ and other cloud parameters.
The velocity dispersion and the mass are well correlated (Pearson coefficient of 0.66), as well as $\sigma_v$ and $(\Sigma R)$, with a Pearson coefficient of 0.68.
\medskip

\noindent \textit{Spatial frequencies} - We found that the GMCs identified with \emph{clumpfind} in the northern filaments of Centaurus A have physical properties distributed in the range of the values found for the Milky Way. However, ALMA does not enable us to fully explore this. As an interferometer, ALMA only recovered the spatial frequencies between $1.3''$ and $14''$ (from 18.1 to $260\: pc$). It is now essential to add observations at higher resolution to reach cloud radii smaller than $10\: pc$ and short-spacing observations for radii larger than 260 pc. We plan to revisit the Larson relations for the large structures in a forthcoming paper in which we will combine ALMA data with our recent ACA observations.
%\textbf{Independently, we also computed (not shown here) the Larson relations for the large structures extracted by the method of \cite{MAMD_2017a} alone. In that case, we found a relation $M\propto R^{2.3}$, consistent with that for the molecular clouds of the Milky Way. However, the velocity dispersions are then much larger, resulting in deviations to the Larson relations. This second extraction method is more sensitive to large structures indentification than \emph{clumpfind} (that tends to bias the extraction towards structures smaller than a few times the beam, i.e. 40 pc). Those large structures are likely to have larger velocity dispersion, but are also more sensitive to short-spacing filtering (that affects the mass and size estimate). We plan to study the Larson relations for the large structures in a forthcoming paper where we will combine ALMA data with our recent ACA observations.}

   \subsection{Excitation of the clouds}

\begin{figure*}[h!]
\centering
  \includegraphics[height=9cm,trim=115 35 244 85,clip=true]{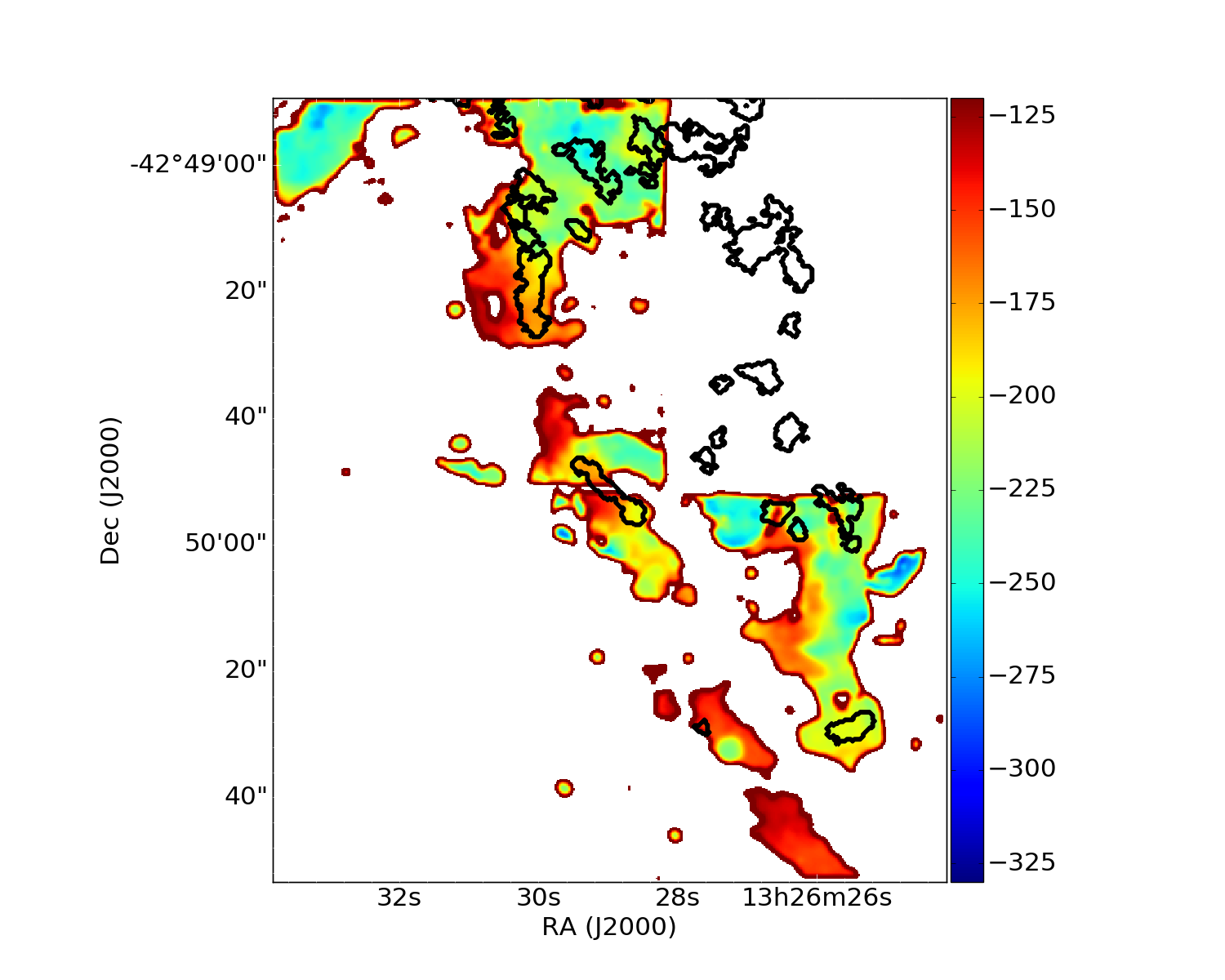}
  \hspace{5mm}
  \includegraphics[height=9cm,trim=240 35 150 85,clip=true]{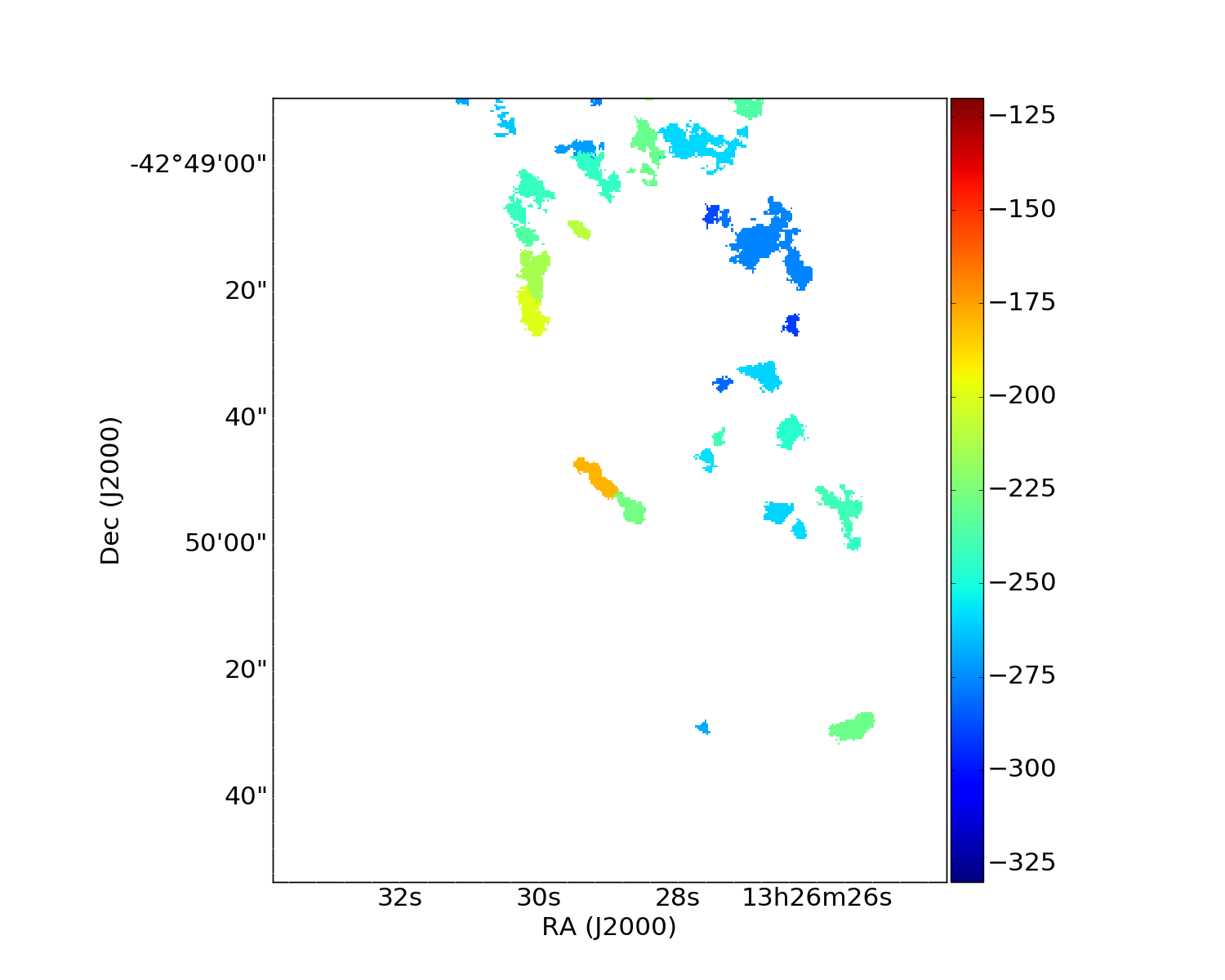}
  \caption{\label{MUSE} Velocity maps in $km.s^{-1}$ of the $H\alpha$ (\emph{left}) and CO emission (\emph{right}). The black contours in the left panel represent the ALMA CO(1-0) emission distribution. The colour scale is the same for both maps, with a velocity range $-330<v<-120\: km.s^{-1}$, relative to Centaurus A. It is thus clear that the CO is blueshifted compared to the $H\alpha$ emission.}
\end{figure*}

   In this section, we focus on the fields of view previously observed in the northern filaments with MUSE \citep{Santoro_2015b}. Observations include the principal excitation optical lines $H\alpha$ $\lambda 6562.8$, [N\rmnum{2}] $\lambda 6583$, $H\beta$ $\lambda 4861.3$, [O\rmnum{3}] $\lambda 4959,5007$, [O\rmnum{1}] $\lambda 6366,$ and the two [S\rmnum{2}] $\lambda 6716,6731$ lines. Figure \ref{MUSE} shows the velocity map of the $H\alpha$ and CO emission, with the same colour scale. By simply comparing the colour of the maps at the location of the CO structures, we clearly found that CO is blueshifted with respect to $H\alpha$. A pixel-by-pixel comparison shows that the velocity CO is on average blueshifted by $v_{CO}-v_{H\alpha}=-35.3\pm 23.2\: km.s^{-1}$.

\begin{figure}[h]
  \centering
  \includegraphics[width=0.75\linewidth,trim=0 0 229 0,clip=true]{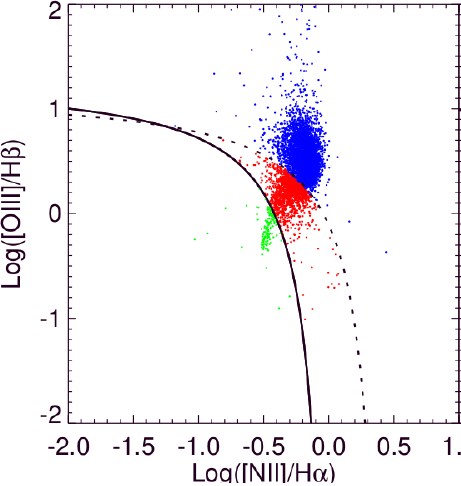}
  \caption{\label{BPT} Pixel-by-pixel BPT diagrams of the CO clumps with MUSE. The black line represents the empirical separation of star formation (green) and AGN/shock-ionised regions (blue). The dotted line shows the extreme upper limit for star formation \citep{Kewley_2006}. The red points correspond to the composite regime.}
\end{figure}

\begin{figure*}[h!]
\centering
  \includegraphics[height=9cm,trim=115 35 150 85,clip=true]{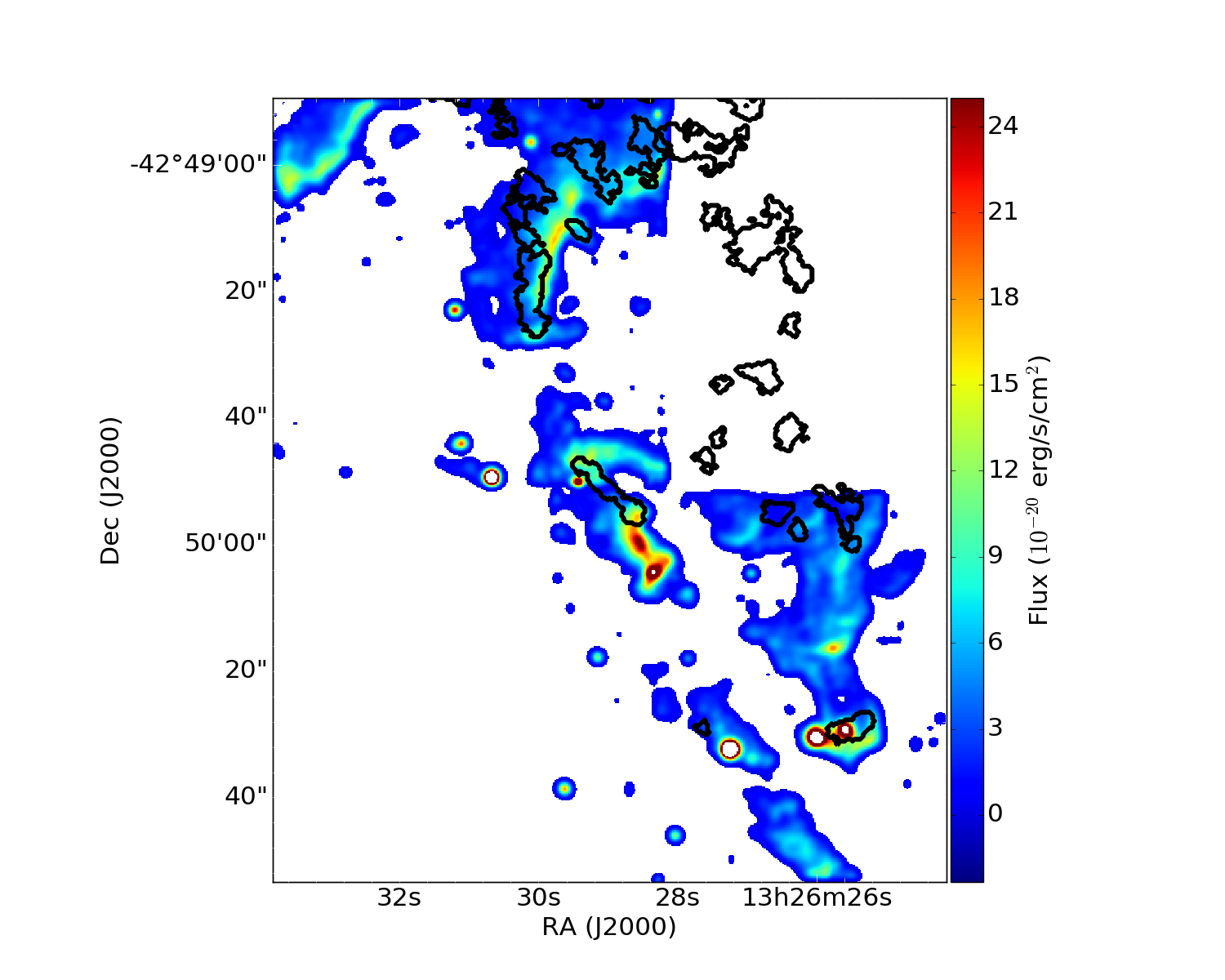}
  \hspace{5mm}
  \includegraphics[height=9cm,trim=255 35 225 85,clip=true]{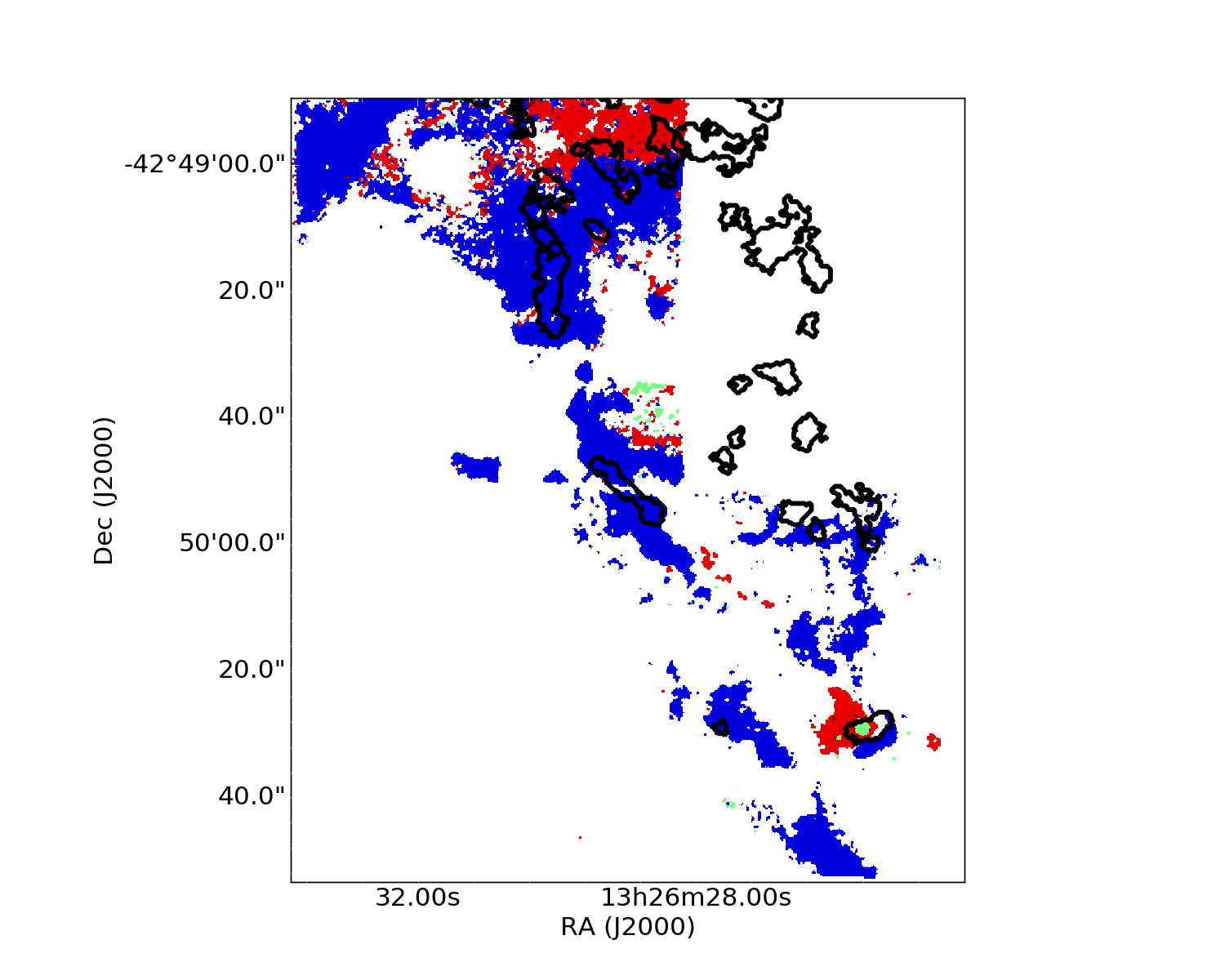}
  \caption{\label{MUSE-BPT} Maps of the $H\alpha$-[N\rmnum{2}] flux from MUSE (\emph{left}) and the excitation processes of the CO structures (\emph{right}). The black contours represent the ALMA CO(1-0) emission distribution. The BPT diagram was computed for the velocity range $-330<v<-120\: km.s^{-1}$. Star formation is represented in green and AGN or shocks are indicated in blue. Red corresponds to a composite of star formation and AGN/shocks. The CO clouds in the Horseshoe complex are associated with shock-excited $H\alpha$ emission, except for one clump that is associated with the star-forming region from \cite{Santoro_2016} and young stellar clusters.}
\end{figure*}

   We computed pixel-by-pixel BPT diagrams based on the $H\alpha$, [N\rmnum{2}], $H\beta$, [O\rmnum{3}], [O\rmnum{1}], and [S\rmnum{2}] lines from MUSE (Fig. \ref{BPT}; \citealt{Baldwin_1981,Kewley_2006}). Each line was fitted by a Gaussian at each spatial resolution element (see \citealt{Hamer_2014} for the details) to measure the flux. Figure \ref{MUSE-BPT} shows a map of the different regions regarding the excitation process within the velocity range $-330<v<-120\: km.s^{-1}$. The structures covered by the MUSE field of views are mostly excited by energy injection from the radio jet or shocks. The small inclusion excited by star formation claimed by \cite{Santoro_2016} is spatially coincident with one CO structure.

\section{Discussion}
\label{sec:discussion}

\noindent \textit{Stability of the clouds} - For each GMC extracted from the ALMA data, we derived the free-fall time $t_{ff}$ and the dynamical time $t_{dyn}$ of the clouds. These timescales are defined by
\begin{eqnarray}
  t_{ff}  &=& 4.4\times 10^7\, \cbra{\frac{n}{cm^{-3}}}^{-1/2}\: yr \\
  t_{dyn} &=& 9.8\times 10^5\, \cbra{\frac{R}{pc}} \cbra{\frac{\sigma_v}{km.s^{-1}}}^{-1}\: yr.
\end{eqnarray}
The free-fall time is about twice as long as the dynamical time, i.e. $t_{dyn}\sim (4.4\pm 1.2)\times 10^6\: yr$ and $t_{ff}\sim (8.3\pm 2.3)\times 10^6\: yr$.
Equivalently, this translates into a virial parameter, $\alpha_{vir}=5\sigma_v^2\, R/(GM)\sim t_{ff}^2/t_{dyn}^2$, with an average value of $11.1\pm 6.6$. This suggests that the molecular clouds in the northern filaments of Centaurus A are not gravitationally bounded structures.
We separated the clouds in two groups: (i) the Horseshoe complex, where gas is excited by shocks, and (ii) the star-forming clouds, associated with $H\alpha$, FUV emission and young stellar clusters. The histogram of the virial parameter does not show any bimodality (Fig. \ref{Hist}) however, the virial parameter tends to be smaller in the star-forming clouds than in the Horseshoe complex.
\bigskip

\noindent \textit{Pressure} - The internal pressure of the clouds is given by
\begin{eqnarray}
  P_{int}=\rho_g \sigma_v^2=4.01\times 10^{-14}\, \cbra{\frac{n}{cm^{-3}}} \cbra{\frac{\sigma_v}{km.s^{-1}}}^2\: dyn.cm^{-2} \label{eq:Pint}
,\end{eqnarray}
where $\rho_g$ is the molecular gas volume density and $\sigma_v$ is the velocity dispersion of the clouds. On average, we found an internal pressure $P_{int}=(6.0\pm 4.7)\times 10^{-11}\: dyn.cm^{-2}$.
Based on X-ray emission, \cite{Kraft_2009,Croston_2009} derived ISM pressures of $\sim 10^{-12}\: dyn.cm^{-2}$ around the inner radio lobes and the northern middle lobe. In the X-ray knots observed by \emph{XMM-Newton} in the northern middle lobe, a thermal pressure has been estimated to be of the order of $\sim 10^{-11}\: dyn.cm^{-2}$ \citep{Kraft_2009}.
Based on these values, \cite{Neff_2015a} derived lower limits of the pressure for the radio features observed at 327 MHz with the VLA. For the diffuse radio emission, these authors estimated that the pressure is of the order of $0.5-1\times 10^{-11}\: dyn.cm^{-2}$. They also found similar values for the radio knots related to the X-ray knots from \cite{Kraft_2009}. However, the pressures derived by \cite{Neff_2015a} are parametrised by the {pressure scaling factor} $\eta$.
The internal pressure we derived for the CO clouds in the northern filaments of Centaurus A is larger than that estimated from X-ray and radio emission \citep{Kraft_2009,Croston_2009,Neff_2015a}. Such a difference might be explain by a pressure scaling factor higher that previously estimated ($\eta\sim 10$; \citealt{Neff_2015a}). Moreover, the previous estimates of the pressure are average values over a large region. It is likely that the pressure is locally higher. One example is the inner radio lobes where the radio pressure is one order of magnitude higher than in the outer parts \citep{Neff_2015a}.
\bigskip

\noindent \textit{Energy injection rate} - Following \cite{MAMD_2017a}, we evaluated the energy injection rate in the clouds. It has been seen numerically (see review by \citealt{Hennebelle_2012}) that turbulent energy decays in one dynamical time if it is not maintained. Therefore the energy dissipation rate is simply the kinetic energy divided by the dynamical time. Here we make the assumption that turbulence has reached a steady state where energy injection is balanced by energy dissipation. This is corroborated by \cite{Neff_2015a} who argued that energy is frequently injected into the outer radio lobes on timescales of the order of 10 Myr, comparable to the dynamical time measured here ($\sim 4\: Myr$). In that case the energy injection rate is:\begin{eqnarray}
  \dot{E}_{inj}=-\frac{1}{2} \frac{M\sigma_v^3}{R}.
\end{eqnarray}
We found $\dot{E}_{inj}=(1.1\pm 1.2)\times 10^2\: L_\odot$, which is a value similar to what is observed in the inner part of the Milky Way where stars are forming actively \citep{MAMD_2017a}.
%\textbf{The large energy injection compresses the gas and triggers the formation of molecular gas in the northern filaments \citep{SalomeQ_2016b}. However, the rate of energy injection is large compare to the free-fall time therefore star formation is quenched \citep{SalomeQ_2016a,SalomeQ_2016b}. The northern filaments is an example of a region with molecular gas but no star formation. We conclude that $\Sigma_{H_2}$ is not a good proxy for $\Sigma_{SFR}$ in this object.}
\bigskip

\noindent \textit{General scenario} - The comparison of the physical properties of the molecular clouds in the northern filaments with those seen in the Milky Way is instructive. Even though the molecular clouds in the northern filaments have physical properties, such as size, mass, and velocity dispersion, close to what is seen in the Milky Way, there is a fundamental difference when studied in detail. It appears that the virial parameter is key. In the Horseshoe region, where no star formation is detected, the virial parameter has a rather broad distribution peaking at $\alpha_{vir}=12.0\pm 7.0$ (see Fig. \ref{Hist} and Table \ref{table:stats}).  This distribution is similar to that of molecular clouds seen in the outer part of the Milky Way, where the star formation efficiency is also low. \cite{MAMD_2017a} argued that the formation of molecular clouds in the outer Milky Way could be related to the dynamical action of infalling matter from the Galactic halo and not by self-gravity or stellar feedback in the disc. The molecular clouds in the Horseshoe region could have a similar origin, whereas in this case the dynamical action of the jet would provide the extra pressure to make the gas transits to the cold phase.

   Interestingly, the virial parameter of the star-forming regions of the northern filament (Fig. \ref{Hist} bottom right, green curve) peaks at a lower value around $\alpha_{vir}=8,$ which indicates that self-gravity has a more important dynamical role for these clouds compared to the Horseshoe region. On the other hand, $\alpha_{vir}$ is still higher on average than in the inner Milky Way, where it peaks at values between 4 and 5 \citep{MAMD_2017a}. This could explain why the star formation efficiency in the northern filaments of Centaurus A is smaller than in the Milky Way.

   These results indicate that energy injected in a system by an external source (infall or AGN feedback for instance) can trigger the formation of molecular gas and formation of stars, but at a level lower than that seen in disc galaxies. This suggests that the Schmidt-Kennicutt relation would only apply to a self-regulated system in which self-gravity and stellar feedback are in balance.
In other situations, the surface density of molecular gas alone is not a good proxy for the star formation rate. Finally, the presence of FUV emission indicates the presence of a population of $\sim 100\: Myr$ stars and suggests different evolutionary stages for the Horseshoe complex and the Vertical filament. These differences could be due to the history of the local kinetic energy injection by the expanding radio jet. However, it is still possible that stars are forming in the Horseshoe complex if they are obscured by dust.

\section{Conclusion}
\label{sec:conclusion}

   In this article, we have presented observations of the $^{12}CO$(1-0) in the northern filaments of Centaurus A at high resolution with the Atacama Large Millimeter/submillimeter Array (ALMA). The region mapped with ALMA corresponds to the region observed with APEX by \cite{SalomeQ_2016b}. The resolution of ALMA ($1.3''\sim 23.8\: pc$) reveals the clumpy structure of the molecular gas, as previously predicted by \cite{SalomeQ_2016a}. However, the present data recover only 20\% of the total molecular gas mass found by APEX. Such a difference is likely due to spatial filtering by the interferometer. Recent observations with the ACA reveals that most of the molecular gas is distributed in more extended structures (Salom\'e et al., in prep.).

   We used the Gaussian decomposition and clustering methods developed by \cite{MAMD_2017a} to extract the signal from the data. The CO emission follows the morphology of the $H\alpha$ emission. We identified two structures in the northern filaments. First, the Horseshoe complex in the bright CO region discovered by \cite{SalomeQ_2016b} is associated with $H\alpha$ emission but not with FUV emission and young stellar clusters. A pixel-by-pixel BPT diagram with optical emission lines from MUSE shows that the CO clouds covered by the MUSE field of views are mostly excited by energy injection from the radio jet or shocks. Second, the Vertical filament is located at the edge of the H\rmnum{1} cloud. Aligned with $H\alpha$, FUV emission, and young stellar clusters, the Vertical filament is likely a region of star formation.

   Applying the \emph{clumpfind} algorithm on the result of the Gaussian decomposition, we extracted 140 molecular clouds. These clouds have size, velocity dispersion, and mass of the same order than molecular clouds in the Milky Way. The range of radius available with the present ALMA observations (less than one order of magnitude) does not enable us to investigate whether the clouds follow the Larson relation or not.

   We derived an estimate of the internal pressure of the clouds. On average, we found an internal pressure of $(6.0\pm 4.7)\times 10^{-11}\: dyn.cm^{-2}$.  This is about one order of magnitude higher that what was derived using X-ray \citep{Kraft_2009,Croston_2009} and radio emission \citep{Neff_2015a}. However, the pressures derived by these authors are average values over a large region and pressure is likely higher locally.

   Finally, we found that the free-fall time is about twice the dynamical time. This indicates that the molecular clouds are not gravitationally unstable. The derived rate of kinetic energy injected in molecular clouds is similar to the typical value found in the inner Milky Way \citep{MAMD_2017a}. However, the star formation rate in the filaments of Centaurus A is much lower than that in the Milky Way. This suggests that, while the energy injected by the jet-gas interaction triggers the HI-to-$H_2$ phase transition, it is high enough to limit gravitational collapse, especially in the Horseshoe complex where no star formation is observed.

\begin{acknowledgements}
   We thank the referee for his/her comments. We thank Edwige Chapillon and the IRAM ARC node for their help during the Phase 2, and Marina Rejkuba for providing the catalogue of stellar clusters. \\

   ALMA is a partnership of ESO (representing its member states), NSF (USA) and NINS (Japan), together with NRC (Canada), NSC and ASIAA (Taiwan), and KASI (Republic of Korea), in cooperation with the Republic of Chile. The Joint ALMA Observatory is operated by ESO, AUI/NRAO and NAOJ. \\

   This work was supported by the ANR grant LYRICS (ANR-16-CE31-0011).
\end{acknowledgements}

\bibliography{Biblio,Biblio_arXiv}

\begin{thebibliography}{57}
\expandafter\ifx\csname natexlab\endcsname\relax\def\natexlab#1{#1}\fi

\bibitem[{Auld {et~al.}(2012)Auld, Smith, Bendo, Pohlen, Wilson, Gomez,
  Cortese, Morganti, Baes, Boselli, Cooray, Davies, Eales, Elbaz, Galametz,
  Isaak, Oosterloo, Page, Rigby, Spinoglio, \& Struve}]{Auld_2012}
Auld, R., Smith, M. W.~L., Bendo, G., {et~al.} 2012, {MNRAS}, 420, 1882

\bibitem[{Baldwin {et~al.}(1981)Baldwin, Phillips, \& Terlevich}]{Baldwin_1981}
Baldwin, J.~A., Phillips, M.~M., \& Terlevich, R. 1981, {PASP}, 93, 5

\bibitem[{Begelman \& Cioffi(1989)}]{Begelman_1989}
Begelman, M.~C. \& Cioffi, D.~F. 1989, {ApJ}, 345, L21

\bibitem[{Bieri {et~al.}(2016)Bieri, Dubois, Silk, Mamon, \&
  Gaibler}]{Bieri_2016}
Bieri, R., Dubois, Y., Silk, J., Mamon, G.~A., \& Gaibler, V. 2016, {MNRAS},
  455, 4166

\bibitem[{Blanco {et~al.}(1975)Blanco, Graham, Lasker, \& Osmer}]{Blanco_1975}
Blanco, V.~M., Graham, J.~A., Lasker, B.~M., \& Osmer, P.~S. 1975, {ApJ}, 198,
  L63

\bibitem[{Bolatto {et~al.}(2013)Bolatto, Wolfire, \& Leroy}]{Bolatto_2013}
Bolatto, A.~D., Wolfire, M., \& Leroy, A.~K. 2013, {ARA\&A}, 51, 207

\bibitem[{Bower {et~al.}(2006)Bower, Benson, Malbon, Helly, Frenk, Baugh, Cole,
  \& Lacey}]{Bower_2006}
Bower, R.~G., Benson, A.~J., Malbon, R., {et~al.} 2006, {MNRAS}, 370, 645

\bibitem[{Charmandaris {et~al.}(2000)Charmandaris, Combes, \& van~der
  Hulst}]{Charmandaris_2000}
Charmandaris, V., Combes, F., \& van~der Hulst, J.~M. 2000, {A\&A}, 356, L1

\bibitem[{Combes(2015)}]{Combes_2015a}
Combes, F. 2015, {IAUS}, 309, 182

\bibitem[{Croston {et~al.}(2009)Croston, Kraft, Hardcastle, Birkinshaw,
  Worrall, Nulsen, Penna, Sivakoff, Jord?n, Brassington, Evans, Forman,
  Gilfanov, Goodger, Harris, Jones, Juett, Murray, Raychaudhury, Sarazin, Voss,
  \& Woodley}]{Croston_2009}
Croston, J.~H., Kraft, R.~P., Hardcastle, M.~J., {et~al.} 2009, {MNRAS}, 395,
  1999

\bibitem[{Croton {et~al.}(2006)Croton, Springel, White, De~Lucia, Frenk, Gao,
  Jenkins, Kauffmann, Navarro, \& Yoshida}]{Croton_2006}
Croton, D.~J., Springel, V., White, S. D.~M., {et~al.} 2006, {MNRAS}, 365, 11

\bibitem[{de~Young(1989)}]{deYoung_1989}
de~Young, D.~S. 1989, {ApJ}, 342, L59

\bibitem[{Dubois {et~al.}(2013)Dubois, Gavazzi, Peirani, \& Silk}]{Dubois_2013}
Dubois, Y., Gavazzi, R., Peirani, S., \& Silk, J. 2013, {MNRAS}, 433, 3297

\bibitem[{Elbaz {et~al.}(2009)Elbaz, Jahnke, Pantin, Le~Borgne, \&
  Letawe}]{Elbaz_2009}
Elbaz, D., Jahnke, K., Pantin, E., Le~Borgne, D., \& Letawe, G. 2009, {A\&A},
  507, 1359

\bibitem[{Emonts {et~al.}(2014)Emonts, Norris, Feain, Mao, Ekers, Miley,
  Seymour, Rottgering, {Villar-Martin}, Sadler, Carilli, Mahony, de~Breuck,
  Stroe, Pentericci, van Moorsel, Drouart, Ivison, Greve, Humphrey, Wylezalek,
  \& Tadhunter}]{Emonts_2014}
Emonts, B. H.~C., Norris, R.~P., Feain, I., {et~al.} 2014, {MNRAS}, 438, 2898

\bibitem[{Feain {et~al.}(2007)Feain, Papadopoulos, Ekers, \&
  Middelberg}]{Feain_2007}
Feain, I.~J., Papadopoulos, P.~P., Ekers, R.~D., \& Middelberg, E. 2007, {ApJ},
  662, 872

\bibitem[{{Fragile} {et~al.}(2017){Fragile}, {Anninos}, {Croft}, {Lacy}, \&
  {Witry}}]{Fragile_arXiv}
{Fragile}, P.~C., {Anninos}, P., {Croft}, S., {Lacy}, M., \& {Witry}, J.~W.~L.
  2017, arXiv:1701.00024

\bibitem[{Gaibler {et~al.}(2012)Gaibler, Khochfar, Krause, \&
  Silk}]{Gaibler_2012}
Gaibler, V., Khochfar, S., Krause, M., \& Silk, J. 2012, {MNRAS}, 425, 438

\bibitem[{Graham \& Price(1981)}]{Graham_1981}
Graham, J.~A. \& Price, R.~M. 1981, {ApJ}, 247, 813

\bibitem[{Hamer {et~al.}(2015)Hamer, Salomé, Combes, \& Salomé}]{Hamer_2015}
Hamer, S., Salomé, P., Combes, F., \& Salomé, Q. 2015, {A\&A}, 575, L3

\bibitem[{Hamer {et~al.}(2014)Hamer, Edge, Swinbank, Oonk, Mittal, {McNamara},
  Russell, Bremer, Combes, Fabian, Nesvadba, {O'Dea}, Baum, Salome, Tremblay,
  Donahue, Ferland, \& Sarazin}]{Hamer_2014}
Hamer, S.~L., Edge, A.~C., Swinbank, A.~M., {et~al.} 2014, {MNRAS}, 437, 862

\bibitem[{Harris {et~al.}(2010)Harris, Rejkuba, \& Harris}]{Harris_2010}
Harris, G. L.~H., Rejkuba, M., \& Harris, W.~E. 2010, {PASA}, 27, 457

\bibitem[{Harrison {et~al.}(2012)Harrison, Alexander, Mullaney, Altieri, Coia,
  Charmandaris, Daddi, Dannerbauer, Dasyra, Del~Moro, Dickinson, Hickox,
  Ivison, Kartaltepe, Le~Floc'h, Leiton, Magnelli, Popesso, Rovilos, Rosario,
  \& Swinbank}]{Harrison_2012}
Harrison, C.~M., Alexander, D.~M., Mullaney, J.~R., {et~al.} 2012, {ApJ}, 760,
  L15

\bibitem[{Hennebelle \& Falgarone(2012)}]{Hennebelle_2012}
Hennebelle, P. \& Falgarone, E. 2012, {A\&AR}, 20, 55

\bibitem[{Heyer {et~al.}(2001)Heyer, Carpenter, \& Snell}]{Heyer_2001}
Heyer, M.~H., Carpenter, J.~M., \& Snell, R.~L. 2001, {ApJ}, 551, 852

\bibitem[{Hughes {et~al.}(2013)Hughes, Meidt, Colombo, Schinnerer, Pety, Leroy,
  Dobbs, {García-Burillo}, Thompson, Dumas, Schuster, \& Kramer}]{Hughes_2013}
Hughes, A., Meidt, S.~E., Colombo, D., {et~al.} 2013, {ApJ}, 779, 46

\bibitem[{Inskip {et~al.}(2008)Inskip, {Villar-Martín}, Tadhunter, Morganti,
  Holt, \& Dicken}]{Inskip_2008}
Inskip, K.~J., {Villar-Martín}, M., Tadhunter, C.~N., {et~al.} 2008, {MNRAS},
  386, 1797

\bibitem[{Israel(1998)}]{Israel_1998}
Israel, F. 1998, {A\&AR}, 8, 237

\bibitem[{Kewley {et~al.}(2006)Kewley, Groves, Kauffmann, \&
  Heckman}]{Kewley_2006}
Kewley, L.~J., Groves, B., Kauffmann, G., \& Heckman, T. 2006, {MNRAS}, 372,
  961

\bibitem[{Klamer {et~al.}(2004)Klamer, Ekers, Sadler, \&
  Hunstead}]{Klamer_2004}
Klamer, I.~J., Ekers, R.~D., Sadler, E.~M., \& Hunstead, R.~W. 2004, {ApJ},
  612, L97

\bibitem[{Kraft {et~al.}(2009)Kraft, Forman, Hardcastle, Birkinshaw, Croston,
  Jones, Nulsen, Worrall, \& Murray}]{Kraft_2009}
Kraft, R.~P., Forman, W.~R., Hardcastle, M.~J., {et~al.} 2009, {ApJ}, 698, 2036

\bibitem[{Kramer {et~al.}(1998)Kramer, Stutzki, Rohrig, \&
  Corneliussen}]{Kramer_1998}
Kramer, C., Stutzki, J., Rohrig, R., \& Corneliussen, U. 1998, {A\&A}, 329, 249

\bibitem[{Kritsuk {et~al.}(2007)Kritsuk, Norman, Padoan, \&
  Wagner}]{Kritsuk_2007}
Kritsuk, A.~G., Norman, M.~L., Padoan, P., \& Wagner, R. 2007, {ApJ}, 665, 416

\bibitem[{Kritsuk {et~al.}(2011)Kritsuk, Norman, \& Wagner}]{Kritsuk_2011}
Kritsuk, A.~G., Norman, M.~L., \& Wagner, R. 2011, {ApJ}, 727, L20

\bibitem[{Marshall {et~al.}(2009)Marshall, Joncas, \& Jones}]{Marshall_2009}
Marshall, D.~J., Joncas, G., \& Jones, A.~P. 2009, {ApJ}, 706, 727

\bibitem[{{McCarthy} {et~al.}(1987){McCarthy}, van Breugel, Spinrad, \&
  Djorgovski}]{McCarthy_1987}
{McCarthy}, P.~J., van Breugel, W., Spinrad, H., \& Djorgovski, S. 1987, {ApJ},
  321, L29

\bibitem[{Miley \& de~Breuck(2008)}]{Miley_2008}
Miley, G. \& de~Breuck, C. 2008, {A\&AR}, 15, 67

\bibitem[{{Miville-Deschênes} {et~al.}(2017){Miville-Deschênes}, Murray, \&
  Lee}]{MAMD_2017a}
{Miville-Deschênes}, M., Murray, N., \& Lee, E.~J. 2017, {ApJ}, 834, 57

\bibitem[{Morganti {et~al.}(1991)Morganti, Robinson, Fosbury,
  di~Serego~Alighieri, Tadhunter, \& Malin}]{Morganti_1991}
Morganti, R., Robinson, A., Fosbury, R. A.~E., {et~al.} 1991, {MNRAS}, 249, 91

\bibitem[{Mould {et~al.}(2000)Mould, Ridgewell, Gallagher~{III}, Bessell,
  Keller, Calzetti, Clarke, Trauger, Grillmair, Ballester, Burrows, Krist,
  Crisp, Evans, Griffiths, Hester, Hoessel, Holtzman, Scowen, Stapelfeldt,
  Sahai, Watson, \& Meadows}]{Mould_2000}
Mould, J.~R., Ridgewell, A., Gallagher~{III}, J.~S., {et~al.} 2000, {ApJ}, 536,
  266

\bibitem[{Neff {et~al.}(2015{\natexlab{a}})Neff, Eilek, \& Owen}]{Neff_2015b}
Neff, S.~G., Eilek, J.~A., \& Owen, F.~N. 2015{\natexlab{a}}, {ApJ}, 802, 88

\bibitem[{Neff {et~al.}(2015{\natexlab{b}})Neff, Eilek, \& Owen}]{Neff_2015a}
Neff, S.~G., Eilek, J.~A., \& Owen, F.~N. 2015{\natexlab{b}}, {ApJ}, 802, 87

\bibitem[{Rees(1989)}]{Rees_1989}
Rees, M.~J. 1989, {MNRAS}, 239, 1

\bibitem[{Reines {et~al.}(2011)Reines, Sivakoff, Johnson, \&
  Brogan}]{Reines_2011}
Reines, A.~E., Sivakoff, G.~R., Johnson, K.~E., \& Brogan, C.~L. 2011, Nature,
  470, 66

\bibitem[{Rejkuba {et~al.}(2001)Rejkuba, Minniti, Silva, \&
  Bedding}]{Rejkuba_2001}
Rejkuba, M., Minniti, D., Silva, D.~R., \& Bedding, T.~R. 2001, {A\&A}, 379,
  781

\bibitem[{Salomé {et~al.}(2016{\natexlab{a}})Salomé, Salomé, Combes, \&
  Hamer}]{SalomeQ_2016b}
Salomé, Q., Salomé, P., Combes, F., \& Hamer, S. 2016{\natexlab{a}}, {A\&A},
  595, A65

\bibitem[{Salomé {et~al.}(2016{\natexlab{b}})Salomé, Salomé, Combes, Hamer,
  \& Heywood}]{SalomeQ_2016a}
Salomé, Q., Salomé, P., Combes, F., Hamer, S., \& Heywood, I.
  2016{\natexlab{b}}, {A\&A}, 586, A45

\bibitem[{Santoro {et~al.}(2015{\natexlab{a}})Santoro, Oonk, Morganti, \&
  Oosterloo}]{Santoro_2015a}
Santoro, F., Oonk, J. B.~R., Morganti, R., \& Oosterloo, T. 2015{\natexlab{a}},
  {A\&A}, 574, A89

\bibitem[{Santoro {et~al.}(2016)Santoro, Oonk, Morganti, Oosterloo, \&
  Tadhunter}]{Santoro_2016}
Santoro, F., Oonk, J. B.~R., Morganti, R., Oosterloo, T.~A., \& Tadhunter, C.
  2016, {A\&A}, 590, A37

\bibitem[{Santoro {et~al.}(2015{\natexlab{b}})Santoro, Oonk, Morganti,
  Oosterloo, \& Tremblay}]{Santoro_2015b}
Santoro, F., Oonk, J. B.~R., Morganti, R., Oosterloo, T.~A., \& Tremblay, G.
  2015{\natexlab{b}}, {A\&A}, 575, L4

\bibitem[{Schiminovich {et~al.}(1994)Schiminovich, van Gorkom, van~der Hulst,
  \& Kasow}]{Schiminovich_1994}
Schiminovich, D., van Gorkom, J.~H., van~der Hulst, J.~M., \& Kasow, S. 1994,
  {ApJ}, 423, L101

\bibitem[{Solomon {et~al.}(1987)Solomon, Rivolo, Barrett, \&
  Yahil}]{Solomon_1987}
Solomon, P.~M., Rivolo, A.~R., Barrett, J., \& Yahil, A. 1987, {ApJ}, 319, 730

\bibitem[{van Breugel {et~al.}(2004)van Breugel, Fragile, Anninos, \&
  Murray}]{vanBreugel_2004}
van Breugel, W., Fragile, C., Anninos, P., \& Murray, S. 2004, {IAUS}, 217, 472

\bibitem[{Wagner {et~al.}(2012)Wagner, Bicknell, \& Umemura}]{Wagner_2012}
Wagner, A.~Y., Bicknell, G.~V., \& Umemura, M. 2012, {ApJ}, 757, 136

\bibitem[{Werner {et~al.}(2014)Werner, Oonk, Sun, Nulsen, Allen, Canning,
  Simionescu, Hoffer, Connor, Donahue, Edge, Fabian, von~der Linden, Reynolds,
  \& Ruszkowski}]{Werner_2014}
Werner, N., Oonk, J. B.~R., Sun, M., {et~al.} 2014, {MNRAS}, 439, 2291

\bibitem[{Williams {et~al.}(1994)Williams, de~Geus, \& Blitz}]{Clumpfind}
Williams, J.~P., de~Geus, E.~J., \& Blitz, L. 1994, {ApJ}, 428, 693

\bibitem[{Zinn {et~al.}(2013)Zinn, Middelberg, Norris, \& Dettmar}]{Zinn_2013}
Zinn, P., Middelberg, E., Norris, R.~P., \& Dettmar, R. 2013, {ApJ}, 774, 66

\end{thebibliography}
\bibliographystyle{aa}

\appendix
%\onecolumn

\section{Properties of molecular clouds}

\begin{table*}[h]
  \centering
  \small
  \caption{\label{table:specCO} Properties of some molecular clouds extracted with the method presented in section \ref{sec:extraction}. The velocities are relative to Cen A ($v_{LSR}\sim 545\: km.s^{-1}$). A catalogue of all the GMCs is available on-line.}
  \begin{tabular}{lcccccc}
    \hline \hline
    Cloud  &   R   & $S_{CO}\Delta v$  &  $v_{cent}$   &  $\Delta v$   &     $M_{H_2}$     &   $\Sigma_{H_2}$    \\
           &  (pc) & ($Jy.km.s^{-1}$)  & ($km.s^{-1}$) & ($km.s^{-1}$) &    ($M_\odot$)    & ($M_\odot.pc^{-2}$) \\ \hline
      1    & 33.39 &       2.909       &    -180.97    &     16.65     & $1.90\times 10^6$ &       425.08        \\
      2    & 29.31 &       2.338       &    -275.31    &     11.01     & $1.52\times 10^6$ &       443.50        \\
      3    & 24.55 &       1.591       &    -228.34    &     11.58     & $1.04\times 10^6$ &       430.25        \\
      4    & 32.94 &       3.579       &    -227.82    &     20.20     & $2.33\times 10^6$ &       537.56        \\
      5    & 26.90 &       1.841       &    -186.42    &     14.60     & $1.20\times 10^6$ &       414.79        \\
      6    & 30.44 &       1.907       &    -226.01    &     14.67     & $1.24\times 10^6$ &       335.25        \\
      7    & 38.33 &       3.359       &    -202.54    &     12.97     & $2.19\times 10^6$ &       372.53        \\
      8    & 31.30 &       1.271       &    -206.29    &      5.73     & $8.28\times 10^5$ &       211.42        \\
      9    & 24.70 &       1.101       &    -176.68    &     10.77     & $7.18\times 10^5$ &       293.93        \\
     10    & 19.60 &       0.774       &    -223.42    &     14.97     & $5.04\times 10^5$ &       328.09        \\ \hline
  \end{tabular}
\end{table*}

\end{document}